\documentclass[10pt,journal,compsoc]{IEEEtran}

\usepackage{graphicx}%
\usepackage{multirow}%
\usepackage{amsmath,amssymb,amsfonts}%
\usepackage{amsthm}%
\usepackage{mathrsfs}%
\usepackage{xcolor}%
\usepackage{textcomp}%
\usepackage{manyfoot}%
\usepackage{booktabs}%
\usepackage{algorithm}%
\usepackage{algorithmicx}%
\usepackage{algpseudocode}%
\usepackage{listings}%
\usepackage{tabularx} % Include this in your preamble
\usepackage{array} % For vertical alignment in tables
\newcolumntype{C}[1]{>{\centering\arraybackslash}m{#1}}
\usepackage{hyperref}
\usepackage{wrapfig}
\usepackage{makecell}
\usepackage{threeparttable}
\usepackage{float}
\usepackage{amsfonts}
\usepackage[most]{tcolorbox}
\ifCLASSOPTIONcompsoc
  \usepackage[nocompress]{cite}
\else
  \usepackage{cite}
\fi
\ifCLASSINFOpdf
\else
\fi
\begin{document}
%
% paper title
% Titles are generally capitalized except for words such as a, an, and, as,
% at, but, by, for, in, nor, of, on, or, the, to and up, which are usually
% not capitalized unless they are the first or last word of the title.
% Linebreaks \\ can be used within to get better formatting as desired.
% Do not put math or special symbols in the title.
\title{A Survey of Quantum Transformers: Architectures, Challenges and Outlooks}
%
%
% author names and IEEE memberships
% note positions of commas and nonbreaking spaces ( ~ ) LaTeX will not break
% a structure at a ~ so this keeps an author's name from being broken across
% two lines.
% use \thanks{} to gain access to the first footnote area
% a separate \thanks must be used for each paragraph as LaTeX2e's \thanks
% was not built to handle multiple paragraphs
%
%
%\IEEEcompsocitemizethanks is a special \thanks that produces the bulleted
% lists the Computer Society journals use for "first footnote" author
% affiliations. Use \IEEEcompsocthanksitem which works much like \item
% for each affiliation group. When not in compsoc mode,
% \IEEEcompsocitemizethanks becomes like \thanks and
% \IEEEcompsocthanksitem becomes a line break with idention. This
% facilitates dual compilation, although admittedly the differences in the
% desired content of \author between the different types of papers makes a
% one-size-fits-all approach a daunting prospect. For instance, compsoc 
% journal papers have the author affiliations above the "Manuscript
% received ..."  text while in non-compsoc journals this is reversed. Sigh.

\author{Hui~Zhang,
        Qinglin~Zhao*,~\IEEEmembership{Senior Member,~IEEE,}
        Mengchu~Zhou,~\IEEEmembership{Fellow,~IEEE,}
        Li~Feng,
        Dusit~Niyato,~\IEEEmembership{Fellow,~IEEE,}
        Shenggen~Zheng,
        and~Lin~Chen,~\IEEEmembership{Member,~IEEE} % <-this % stops a space    
\IEEEcompsocitemizethanks{\IEEEcompsocthanksitem H. Zhang, Q. Zhao (corresponding author) and L. Feng are with Faculty of Innovation Engineering, Macau University of Science and Technology, Macao 999078, China. \protect\\
% note need leading \protect in front of \\ to get a newline within \thanks as
% \\ is fragile and will error, could use \hfil\break instead.
E-mail: h.zhang2023@hotmail.com; qlzhao@must.edu.mo; lfeng@must.edu.mo
\IEEEcompsocthanksitem M. Zhou is with the Department of Electrical and Computer Engineering, New Jersey Institute of Technology, Newark, NJ 07102, USA.
E-mail: zhou@njit.edu
\IEEEcompsocthanksitem D. Niyato is with the College of Computing and Data Science, Nanyang Technological University, Singapore.
E-mail: dniyato@ntu.edu.sg
\IEEEcompsocthanksitem S. Zheng is with QuantumScience Center of Guangdong-Hong Kong-Macao Greater Bay Area, Shenzhen 518045, China.
E-mail: zhengshenggen@quantumsc.cn
\IEEEcompsocthanksitem L. Chen is with the Engineering Research Centre of Applied Technology on Machine Translation and Artificial Intelligence, Macao Polytechnic University, Macao 999078, China.
E-mail: lchen@mpu.edu.mo}% <-this % stops an unwanted space
\thanks{}}

% note the % following the last \IEEEmembership and also \thanks - 
% these prevent an unwanted space from occurring between the last author name
% and the end of the author line. i.e., if you had this:
% 
% \author{....lastname \thanks{...} \thanks{...} }
%                     ^------------^------------^----Do not want these spaces!
%
% a space would be appended to the last name and could cause every name on that
% line to be shifted left slightly. This is one of those "LaTeX things". For
% instance, "\textbf{A} \textbf{B}" will typeset as "A B" not "AB". To get
% "AB" then you have to do: "\textbf{A}\textbf{B}"
% \thanks is no different in this regard, so shield the last } of each \thanks
% that ends a line with a % and do not let a space in before the next \thanks.
% Spaces after \IEEEmembership other than the last one are OK (and needed) as
% you are supposed to have spaces between the names. For what it is worth,
% this is a minor point as most people would not even notice if the said evil
% space somehow managed to creep in.

% The paper headers
\markboth{}%
{Shell \MakeLowercase{\textit{et al.}}: A Survey of Quantum Transformers: Architectures, Challenges and Outlooks}
% The only time the second header will appear is for the odd numbered pages
% after the title page when using the twoside option.
% 
%  Note that you probably will NOT want to include the author's 
%  name in the headers of peer review papers.                   
% You can use \ifCLASSOPTIONpeerreview for conditional compilation here if
% you desire.

% The publisher's ID mark at the bottom of the page is less important with
% Computer Society journal papers as those publications place the marks
% outside of the main text columns and, therefore, unlike regular IEEE
% journals, the available text space is not reduced by their presence.
% If you want to put a publisher's ID mark on the page you can do it like
% this:
%\IEEEpubid{0000--0000/00\$00.00~\copyright~2015 IEEE}
% or like this to get the Computer Society new two part style.
%\IEEEpubid{\makebox[\columnwidth]{\hfill 0000--0000/00/\$00.00~\copyright~2015 IEEE}%
%\hspace{\columnsep}\makebox[\columnwidth]{Published by the IEEE Computer Society\hfill}}
% Remember, if you use this you must call \IEEEpubidadjcol in the second
% column for its text to clear the IEEEpubid mark (Computer Society jorunal
% papers don't need this extra clearance.)

% use for special paper notices
%\IEEEspecialpapernotice{(Invited Paper)}

% for Computer Society papers, we must declare the abstract and index terms
% PRIOR to the title within the \IEEEtitleabstractindextext IEEEtran
% command as these need to go into the title area created by \maketitle.
% As a general rule, do not put math, special symbols or citations
% in the abstract or keywords.
\IEEEtitleabstractindextext{%
\begin{abstract}
Quantum Transformers integrate the representational power of classical Transformers with the computational advantages of quantum computing, leveraging quantum features such as superposition and entanglement. Since 2022, research in this area has rapidly expanded, giving rise to diverse technical paradigms and early applications. To address the growing need for consolidation, this paper presents the first comprehensive, systematic, and in-depth survey of quantum Transformer models.
First, we delineate the research scope, focusing on improving Transformer parts with quantum methods, and introduce foundational concepts in classical Transformers and quantum machine learning. 
Then, we organize existing studies into two main paradigms: Parameterized Quantum Circuits (PQC)-based and Quantum Linear Algebra (QLA)-based, with PQC-based paradigm further divided into QKV (Query, Key and Value)-only Quantum Mapping, Quantum Pairwise Attention, Quantum Holistic Attention. and Quantum-Assisted Optimization, analyzing their core mechanisms and architectural traits.
We also summarize empirical results that demonstrate preliminary quantum advantages, especially on small-scale tasks or resource-constrained settings. Following this, we examine key technical challenges—such as complexity-resource trade-offs, scalability and generalization limitations, and trainability issues including barren plateaus—and provide potential solutions, including quantumizing classical transformer variants with lower complexity, hybrid designs, and improved optimization strategies.
Finally, we propose several promising future directions, e.g., scaling quantum modules into large architectures, applying quantum Transformers to domains with inherently quantum data (e.g., physics, chemistry), and developing theory-driven designs grounded in quantum information science. 
This survey will help researchers and practitioners quickly grasp the overall landscape of current quantum Transformer research and promote future developments in this emerging field.
\end{abstract}

% Note that keywords are not normally used for peerreview papers.
}

% make the title area
\maketitle

% To allow for easy dual compilation without having to reenter the
% abstract/keywords data, the \IEEEtitleabstractindextext text will
% not be used in maketitle, but will appear (i.e., to be "transported")
% here as \IEEEdisplaynontitleabstractindextext when the compsoc 
% or transmag modes are not selected <OR> if conference mode is selected 
% - because all conference papers position the abstract like regular
% papers do.
\IEEEdisplaynontitleabstractindextext
% \IEEEdisplaynontitleabstractindextext has no effect when using
% compsoc or transmag under a non-conference mode.

% For peer review papers, you can put extra information on the cover
% page as needed:
% \ifCLASSOPTIONpeerreview
% \begin{center} \bfseries EDICS Category: 3-BBND \end{center}
% \fi
%
% For peerreview papers, this IEEEtran command inserts a page break and
% creates the second title. It will be ignored for other modes.
\IEEEpeerreviewmaketitle

\IEEEraisesectionheading{\section{Introduction}\label{sec:introduction}}
% Computer Society journal (but not conference!) papers do something unusual
% with the very first section heading (almost always called "Introduction").
% They place it ABOVE the main text! IEEEtran.cls does not automatically do
% this for you, but you can achieve this effect with the provided
% \IEEEraisesectionheading{} command. Note the need to keep any \label that
% is to refer to the section immediately after \section in the above as
% \IEEEraisesectionheading puts \section within a raised box.

% The very first letter is a 2 line initial drop letter followed
% by the rest of the first word in caps (small caps for compsoc).
% 
% form to use if the first word consists of a single letter:
% \IEEEPARstart{A}{demo} file is ....
% 
% form to use if you need the single drop letter followed by
% normal text (unknown if ever used by the IEEE):
% \IEEEPARstart{A}{}demo file is ....
% 
% Some journals put the first two words in caps:
% \IEEEPARstart{T}{his demo} file is ....
% 
% Here we have the typical use of a "T" for an initial drop letter
% and "HIS" in caps to complete the first word.

Quantum Machine Learning (QML), as an interdisciplinary field at the intersection of quantum computing and classical machine learning, has witnessed rapid advancements in recent years, garnering significant attention from both academia and industry \cite{bib1}. The core objective of QML is to leverage the unique properties of quantum computing—such as superposition, entanglement, and interference—to enhance data processing capabilities and address the computational complexity and efficiency bottlenecks inherent in classical methods\cite{bib2, bib7}. Its potential for enhanced representation learning and computational acceleration has positioned QML as one of the most active areas of research \cite{bib5, bib9, bib12, bib26}.

\begin{tcolorbox}[colback=gray!5, colframe=black, title=List of Abbreviations]
\footnotesize
\begin{tabular}{@{}lp{0.85\textwidth}@{}}
\textbf{QML}  & Quantum Machine Learning \\
\textbf{PQC}  & Parameterized Quantum Circuit \\
\textbf{NISQ} & Noisy Intermediate-Scale Quantum \\
\textbf{QLA}  & Quantum Linear Algebra \\
\textbf{VQA}  & Variational Quantum Algorithm \\
\textbf{QE} & Qubit Encoding \\
\textbf{AE} & Amplitude Encoding \\
\textbf{UE} & Unary Encoding \\
\textbf{DE} & Data Encoding \\
\textbf{HEP}  & High-Energy Physics \\
\textbf{GPQSA}  & Gaussian Projected Quantum Self-Attention \\
\textbf{APDM}  & Amplitude-Phase Decomposed Measurements \\
\textbf{QLS}  & Quantum Logical Similarity \\
\textbf{VQE}  & Variational Quantum Eigensolver \\
\textbf{RBS}  & Rotated Beam Splitter \\
\textbf{QOL}  & Quantum Orthogonal Layer \\
\textbf{QPIXL}  & Quantum pixel representations \\
\textbf{QFT}  & Quantum Fourier Transform \\
\textbf{DSM}  & Doubly Stochastic Matrix \\
\textbf{LCU}  & Linear Combination of Unitaries \\
\textbf{CLCU}  & Complex Linear Combination of Unitaries \\
\textbf{QSVT} & Quantum Singular Value Transformation \\
\textbf{qRAM} & quantum Random Access Memory \\
\textbf{MLP}  & Multilayer Perceptron \\
\textbf{QCT}  & Quantum Compound Transformer
\end{tabular}
\end{tcolorbox}

Among the various directions in QML, one of the most widely explored involves translating classical models into their quantum counterparts. Notable examples include Quantum Convolutional Neural Networks (QCNNs) \cite{bib39, bib18, hur2022qcnn}, Quantum Recurrent Neural Networks (QRNNs) \cite{bausch2020qrnn, bib23}, and Quantum Generative Adversarial Networks (QGANs) \cite{bib24, bib25}, which have been applied to a range of tasks such as classification, sequence modeling, and data generation. However, these models often inherit structural constraints from their classical counterparts. For instance, QCNNs excel at capturing local patterns but are limited in global feature modeling \cite{hur2022qcnn}. QRNNs face challenges in parallelism and scalability for long sequences \cite{bausch2020qrnn}, while QGANs suffer from training instability and lack reliable convergence guarantees on near-term hardware \cite{zoufal2019qgan}. These limitations motivate the exploration of more flexible architectures that can fully leverage the power of quantum computation.

Since its introduction \cite{vaswani2017attention}, the classical Transformer architecture has achieved groundbreaking success in fields such as natural language processing (NLP) and computer vision (CV), primarily due to the efficiency of its self-attention mechanism. For instance, BERT \cite{devlin2018bert} and GPT-4 \cite{achiam2023gpt}, both utilizing the transformer as their backbone, have demonstrated remarkable language understanding capabilities in NLP tasks, while Vision Transformer (ViT) \cite{dosovitskiy2020vit} has challenged the dominance of convolutional neural networks (CNNs) in image processing. However, the computational complexity of the self-attention mechanism grows quadratically with sequence length, resulting in significant computational overhead. This limitation has motivated researchers to explore the integration of quantum mechanism principles with Transformer architectures, giving rise to the emerging research direction of Quantum Transformers.

\begin{figure}[htbp]
    \centering
    \includegraphics[width=0.48\textwidth]{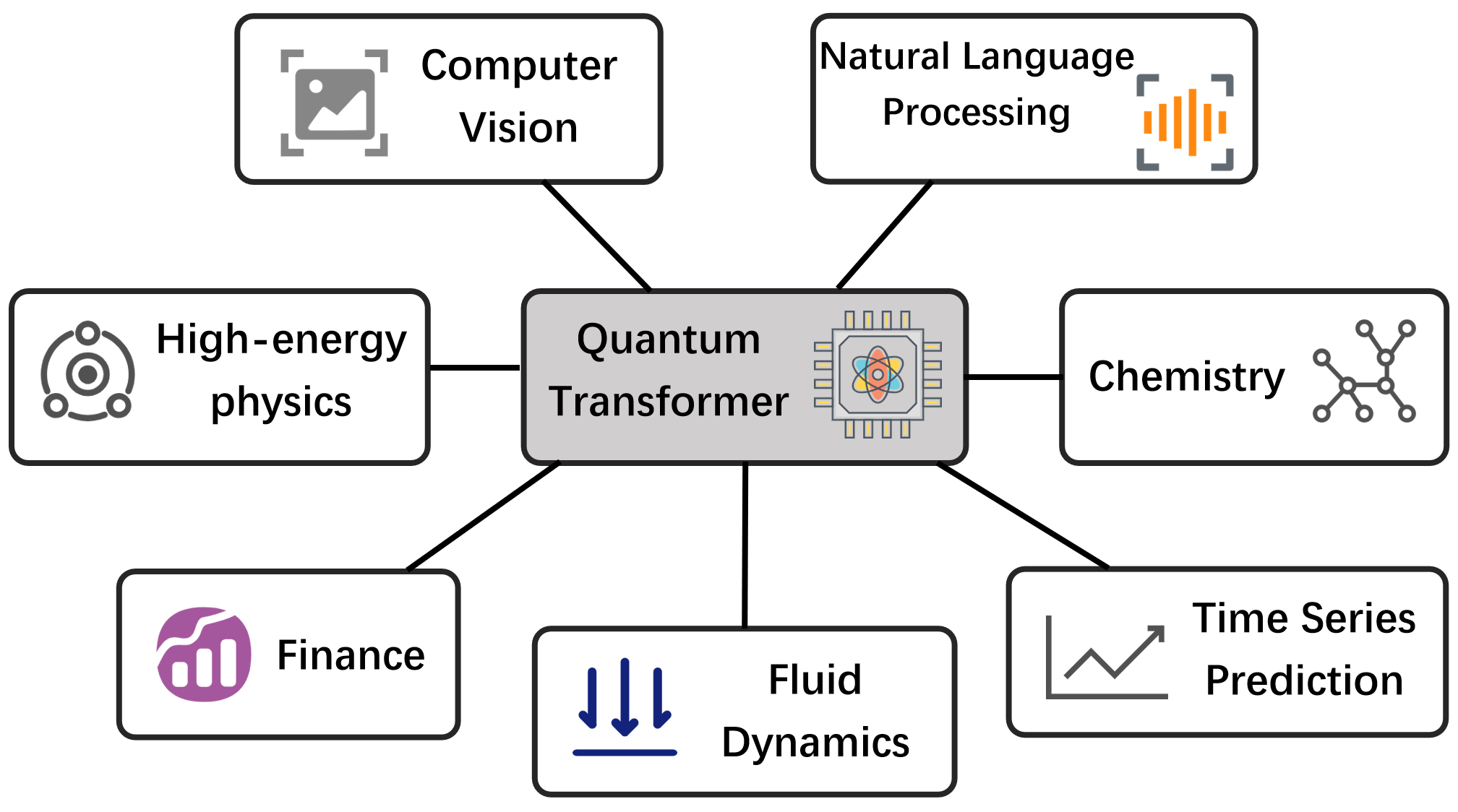}
    \caption{The application scenarios of Quantum Transformers} \label{Application}
\end{figure}

Building on the Transformer’s success and QML’s computational strengths, Quantum Transformers aim to combine global attention mechanisms with quantum-enhanced expressivity while investigating quantum strategies to reduce the quadratic scaling of self-attention. Reflecting this promising direction, recent years have witnessed a rapid increase in research activities focused on the design and application of quantum Transformers. Since 2022 \cite{li2024quantum}, dozens of studies have investigated various quantum adaptations of Transformer architectures \cite{chen2025quantum, zhao2024QKSAN, kerenidis2024quantum, guo2024quantum}, ranging from novel circuit formulations and attention mechanisms to early-stage experimental implementations on quantum hardware and simulators. These studies not only highlight the potential of quantum Transformers but also showcase the diversity of their implementation approaches. Furthermore, as the field progresses, some quantum Transformer models have also begun to find use in practical tasks, including CV, NLP, high-energy physics, quantum chemistry and so on, as summarized in Fig.~\ref{Application}.

\begin{figure*}[htbp]
    \centering
    \includegraphics[width=0.95\textwidth]{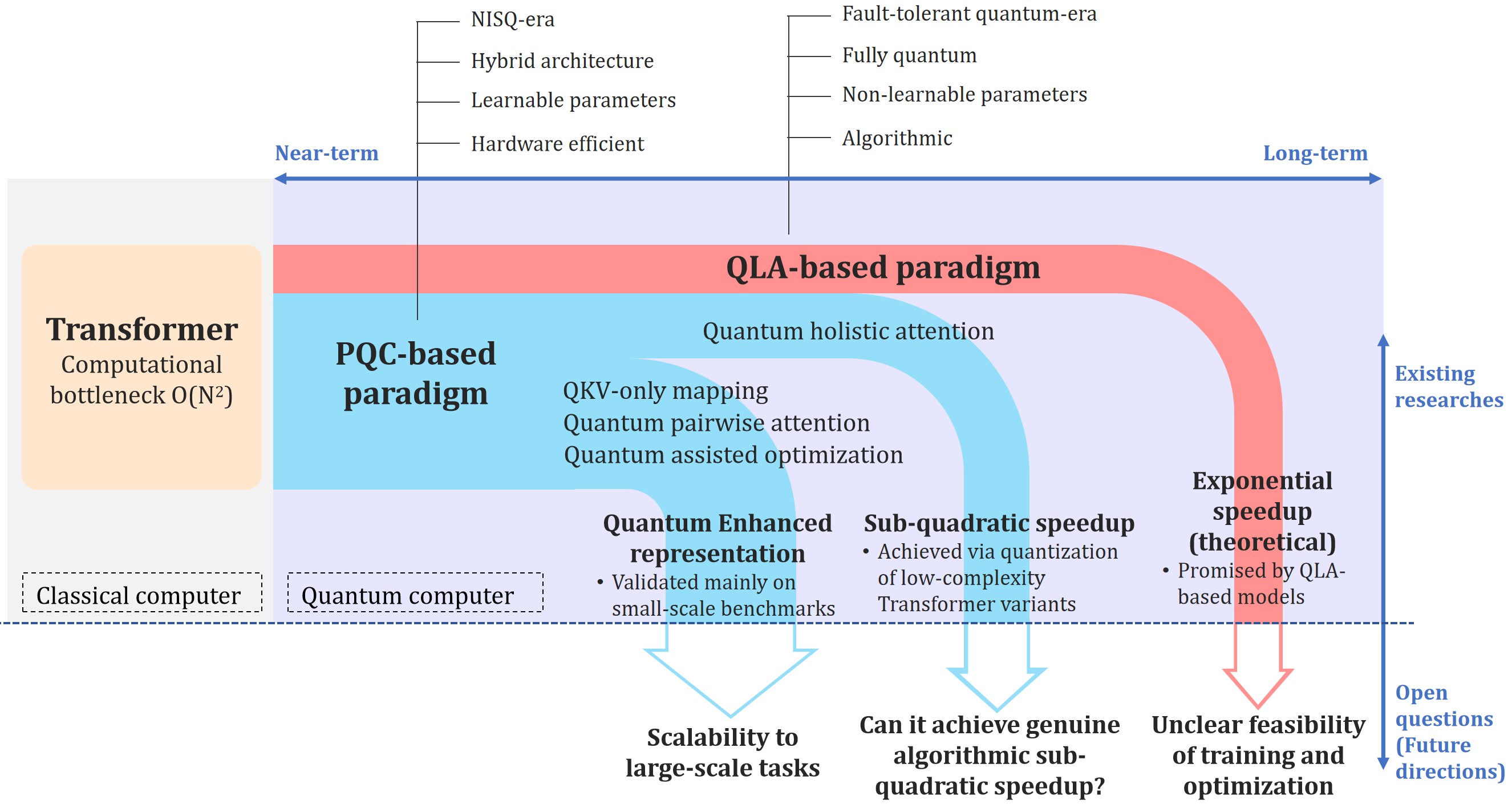}
    \caption{The Strategic Roadmap of Quantum Transformers. The classical Transformer's $O(N^2)$ computational bottleneck has motivated the exploration of quantum solutions. These solutions have diverged into two main paradigms. The near-term, PQC-based path (blue stream) leverages NISQ-era hardware to achieve Quantum Enhanced Representation, leading to open questions about scalability to large-scale tasks and genuine algorithmic sub-quadratic speedup. The long-term, QLA-based path (red stream) targets fault-tolerant hardware with the goal of Theoretical Exponential Speedup, facing fundamental challenges in training and optimization. The figure maps the classification of existing research to this roadmap, highlighting current progress and future directions.} \label{summary}
\end{figure*}

Broadly speaking, current research on quantum Transformers follows two main implementation paradigms, as visually detailed in the strategic roadmap (Fig.~\ref{summary}). 

\textit{1) PQC-based Quantum Transformers.} This approach, representing the near-term research path, leverages parameterized quantum circuits (PQC) to simulate or replace key components of the Transformer \cite{bib4, bib36}. These methods aim to achieve quantum advantages under the constraints of current Noisy Intermediate-Scale Quantum (NISQ) \cite{bharti2022noisy} devices, primarily by enhancing the model's expressibility. To bring clarity to this diverse landscape, we propose a novel subdivision of this paradigm into four distinct sub-directions: some studies focus on a shallow integration by replacing only the QKV mapping (\textit{QKV-only mapping}) \cite{li2024quantum, zhang2024light-weight}; others attempt a deeper quantumization of the similarity calculation (\textit{Quantum pairwise attention}) \cite{chen2025quantum, zhao2024QKSAN}; some reformulate the entire mechanism to capture global dependencies in a single step (\textit{Quantum holistic attention}) \cite{kerenidis2024quantum}; and a few use quantum algorithms to refine the classical attention matrix (\textit{Quantum assisted optimization}). The main achievement of this paradigm to date has been the demonstration of \textit{Quantum Enhanced Representation}, which has been validated on various small-scale benchmarks. However, these progresses leads to critical open questions regarding the \textit{scalability of these methods to large-scale tasks} and whether they can achieve \textit{genuine algorithmic speedup} or merely redistribute computational costs.

\textit{2) QLA-based Quantum Transformers.} This approach, representing the long-term vision, utilizes quantum linear algebra (QLA) techniques, such as block encoding and Quantum Singular Value Transformation (QSVT) \cite{chakraborty2018power, gilyen2019quantum}, to accelerate matrix operations within Transformers. By aiming to significantly reduce the classical computational complexity, QLA-based methods promise a theoretical \textit{Exponential Speedup}. However, their practical implementation relies on high-fidelity quantum operations and large-scale qubit support, requiring fault-tolerant quantum computers. As a result, and as noted in our roadmap, research in this area remains largely theoretical at present, with the \textit{unclear feasibility of training and optimization} methods representing the most significant unresolved challenge.

\subsection{Motivation}

Although quantum Transformer research is attracting increasing attention alongside the emergence of various technical approaches, a comprehensive, systematic review of the existing literature is currently absent. The diversity of these methods and the evolving research trajectories in this field prompt several critical questions: How are these quantum Transformer architectures constructed? What are their differences? Do they demonstrate quantum advantages and if so, under what conditions? Which technical pathways appear the most promising? Consequently synthesizing these research efforts is crucial for providing a clear understanding of the field's landscape and outlining potential future directions.

To offer a comprehensive review of quantum transformers and to answer the above-mentioned questions, this work systematically analyzes and synthesizes the existing quantum Transformer models, evaluating their technical characteristics, advantages, and challenges while providing insights into their future development.

\subsection{Scope and Research Focus}

\begin{table*}[htbp]
\centering
\begingroup
\renewcommand{\arraystretch}{0.95}
\footnotesize
\caption{The Quantum Transformer papers discussed in this review. In the ‘Research Focus’ column, a star ($\bigstar$) indicates an \textit{Original Design}, a square ($\blacksquare$) corresponds to \textit{Method Improvement}, and a triangle ($\blacktriangle$) signifies \textit{Application-Oriented} contributions. In the ‘Application’ column, a dash (—) indicates that the method has not been experimentally validated.}
\label{papers}
\begin{tabular}{C{1.8cm}C{3cm}C{1.5cm}C{4cm}C{1.5cm}C{2cm}}
\toprule%
\textbf{Implementation Paradigm} & \textbf{Authors}  & \textbf{Year} & \textbf{Key Quantum Tools} & \textbf{Research Focus}  & \textbf{Application} \\ \midrule
\multirow{24}{=}{PQC-based Quantum Transformer} & Li et al. \cite{li2024quantum} & 2022 & PQC, GPQSA & $\bigstar$ & NLP \\ \cmidrule{2-6}
& Zhang et al. \cite{zhang2024light-weight} & 2024 & PQC, APDM & $\blacksquare$ & NLP \& CV \\ \cmidrule{2-6}
& Wei et al. \cite{wei2023povm} & 2023 & PQC, POVM & $\blacksquare$ & NLP \\ \cmidrule{2-6}
& Nguyen et al. \cite{bib64} & 2024 & PQC & \large$\blacktriangle$ & CV \\ \cmidrule{2-6}
& Comajoan et al. \cite{cara2024quantum} & 2024 & PQC & \large$\blacktriangle$ & HEP \\ \cmidrule{2-6}
& Unlu et al. \cite{unlu2024hybrid} & 2024 & PQC & \large$\blacktriangle$ & HEP \\ \cmidrule{2-6}
& Chakraborty et al. \cite{chakraborty2025integrating} & 2025 & PQC & \large$\blacktriangle$ & Time Series \\ \cmidrule{2-6}
& Dutta et al. \cite{dutta2024qadqn} & 2024 & PQC & \large$\blacktriangle$ & Finance \\ \cmidrule{2-6}
& Dutta et al. \cite{dutta2024aq} & 2024 & PQC & \large$\blacktriangle$ & Fluid Dynamics \\ \cmidrule{2-6}
& He et al. \cite{he2024training} & 2024 & PQC & \large$\blacktriangle$ & NLP \\ \cmidrule{2-6}
& Chen et al. \cite{chen2025quantum} & 2024 & PQC, swap test, partial trace & $\bigstar$ & NLP \\ \cmidrule{2-6}
& Zhang et al. \cite{zhang2025hqvit} & 2025 & PQC, swap test, conditional measurement & $\bigstar$ & CV \\ \cmidrule{2-6}
& Smaldone et al. \cite{smaldone2025hybrid} & 2025 & PQC, Hadamard test & $\bigstar$ & Chemistry \\ \cmidrule{2-6}
& Zhao et al. \cite{zhao2024QKSAN} & 2024 & PQC, quantum kernel, conditional measurement & $\bigstar$ & CV \\ \cmidrule{2-6}
& Kamata et al. \cite{kamata2025molecular} & 2025 & PQC, quantum kernel, matrix product state & $\bigstar$ & Chemistry \\ \cmidrule{2-6}
& Shi et al. \cite{shi2024QSAN} & 2022 & PQC, QLS & $\bigstar$ & CV \\ \cmidrule{2-6}
& Zheng et al. \cite{zheng2023Design} & 2023 & PQC & $\bigstar$ & NLP \\ \cmidrule{2-6}
& Shi et al. \cite{shi2023natural} & 2023 & PQC & $\bigstar$ & NLP \\ \cmidrule{2-6}
& Chen et al. \cite{chen2025quantum-c} & 2025 & PQC, improved Hadamard test, CLCU & $\bigstar$ & CV \\ \cmidrule{2-6}
& Kerenidis et al. \cite{kerenidis2024quantum} & 2022 & RBS gate, QOL & $\bigstar$ & CV \\ \cmidrule{2-6}
& Alessandro et al. \cite{tesi2024quantum} & 2024 & RBS gate, QOL & \large$\blacktriangle$ & HEP \\ \cmidrule{2-6}
& Evans et al. \cite{evans2024learning} & 2024 & PQC, QFT & $\bigstar$ & -- \\ \cmidrule{2-6}
& Gao et al. \cite{gao2023fast} & 2023 & Grover algorithm & $\bigstar$ & -- \\ \cmidrule{2-6}
& Born et al. \cite{born2025quantum} & 2025 & PQC & $\bigstar$ & CV \\ \midrule
\multirow{4}{=}{QLA-based Quantum Transformer} & Guo et al. \cite{guo2024quantum} & 2024 & QSVT, LCU & $\bigstar$ & -- \\ \cmidrule{2-6}
& Liao et al. \cite{liao2024gpt} & 2024 & Amplitude estimation, phase estimation & $\bigstar$ & -- \\ \cmidrule{2-6}
& Khatri et al. \cite{khatri2024quixer} & 2024 & PQC, LCU, QSVT & $\bigstar$ & NLP \\ \cmidrule{2-6}
& Xue et al. \cite{xue2024end} & 2024 & QSVT, qRAM & \large$\blacktriangle$ & CV \\ 
\bottomrule
\end{tabular}
\endgroup
\end{table*}

This review focuses on QML models that explicitly adopt Transformer-style architectures, excluding those that only use quantum preprocessing/postprocessing without altering Transformer blocks \cite{tariq2024deep,liu2024quantum,yu2024gqwformer,zhang2025quantum}. After careful filtering, we highlight a selection of representative and influential studies, summarized in Table~\ref{papers} by implementation paradigm, authorship, publication year, research focus, and application domain. The field is still young and evolving, with all included works published within the past three years.

Research efforts span three main directions: \textit{Original Design}, \textit{Method Improvement} based on existing architectures, and \textit{Application-Oriented} studies targeting specific domains. These applications cover a broad range of fields, including computer vision, natural language processing, high-energy physics, and chemistry, as illustrated in Fig.~\ref{Application}.

Of the selected studies, 24 adopt the PQC paradigm and 4 follow the QLA approach. We place greater emphasis on PQC-based methods, as they are more compatible with current quantum hardware and offer more practical insights into the near-term potential of quantum Transformers. Their technical roadmaps, experimental results, and application performance are detailed in Fig.\ref{roadmap} and Tables\ref{NLP results}, \ref{CV results}, and \ref{RW results}. QLA-based approaches, while largely theoretical at present, are also discussed as promising directions for the fault-tolerant quantum era.

Although this review does not aim to be exhaustive, it seeks to capture the core trends and most impactful developments in quantum Transformer research.

\subsection{Contributions}

The main contributions of this paper are as follows:

\textit{i) First Comprehensive Review of Quantum Transformer Research.} We present the first comprehensive and in-depth survey of quantum Transformer models, systematically consolidating diverse research efforts. Our review encompasses both PQC-based and QLA-based approaches, offering a holistic perspective on the evolution of architectures, core methodologies, and emerging application domains.

\textit{ii) Systematic Technical and Application-Oriented Analysis.} We conduct a thorough and structured analysis of existing quantum Transformer architectures by introducing a novel classification framework. This framework clarifies the technical underpinnings and algorithmic patterns of different models. In addition, we summarize current application trends and highlight the preliminary advantages demonstrated in early-stage implementations.

\textit{iii) In-depth Analysis of Challenges and Potential Solutions.} We identify and analyze key challenges in the field, with particular emphasis on the trade-offs between model complexity and quantum resource requirements under NISQ constraints. Moreover, we provide initial insights into possible solution directions, including circuit design optimization, hybrid strategies, and more robust encoding schemes.

\textit{iv) Valuable Outlooks.} Drawing on cutting-edge developments and novel insights, we outline several forward-looking and constructive research directions. These perspectives are intended to inspire future theoretical innovations and practical explorations in this rapidly evolving area.

The remainder of this paper is organized as follows: Section \ref{sec2} introduces the classical Transformer mechanism and QML fundamentals relevant to Quantum Transformers. Section \ref{sec3} introduces the technical approaches to building these quantum transformer models. Section \ref{sec4} discusses the preliminary demonstration of quantum advantages and its applications. While section \ref{sec5} presents the key challenges in current quantum transformer studies, Section \ref{sec6} gives several possible future directions to address these challenges. Section \ref{sec7} gives a comprehensive summary and conclusion.

%%%%%%%%%%%%%%%%%%%%%%%%%%%%%%%%%%%%%%%%%%%%%%%%%%%%%%%%%%%%%%%%%%%%%%%%%%%%%%%%%%%%%%%%%%%%%%%%%%%%%%%%%%

\section{Fundamentals}\label{sec2}

Before introducing existing quantum transformer models, it is necessary to understand the fundamentals about Classical Transformer and QML.

\subsection{Classical Transformer}\label{sec2.1}

\begin{figure}[t]
    \centering
    \includegraphics[width=0.45\textwidth]{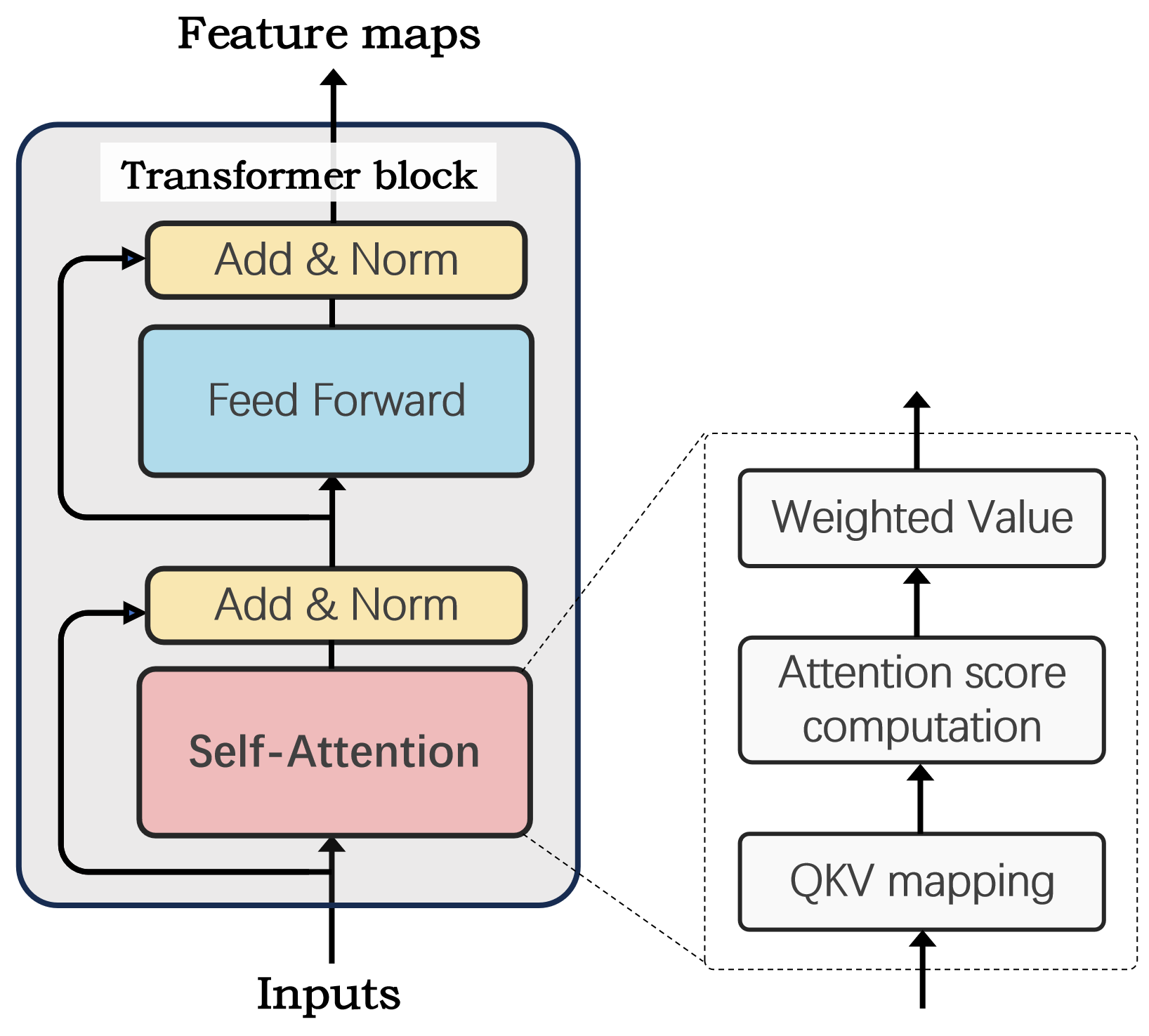}
    \caption{The structure of Transformer block. It processes the input through a self-attention mechanism followed by a feed-forward network. Each sub-layer is wrapped with an Add \& Norm operation. Specifically, the input is first passed through the self-attention layer (with QKV mapping, attention score computation, and value weighting), then normalized and added to the original input. This output is further processed by a feed-forward layer, again followed by Add \& Norm, producing the final feature maps.} \label{transformer}
\end{figure}

The original Transformer follows an Encoder-Decoder structure. In different tasks, it can flexibly use only the Encoder or Decoder. Regardless of the specific usage, both consist of multiple stacked Transformer blocks, each containing the following key components: self-attention mechanisms, multi-head attention, position-wise feedforward networks, and residual connections with normalization. The structure of a Transformer block is shown in Fig. \ref{transformer}.

\textbf{Self-Attention Mechanism.}
Self-attention allows each position in a sequence to attend to all others, capturing long-range dependencies \cite{vaswani2017attention}. It involves three main steps:

\textit{QKV mapping} - Given an input sequence \(\mathbf{X} \in \mathbb{R}^{N \times d}\), where \( N \) is the sequence length and \( d \) is the feature dimension of each token, we project it into three different spaces to obtain queries \((\mathbf{Q})\), keys \((\mathbf{K})\), and values \((\mathbf{V})\):
\[
\mathbf{Q} = \mathbf{W}_Q \mathbf{X}, \quad
\mathbf{K} = \mathbf{W}_K \mathbf{X}, \quad
\mathbf{V} = \mathbf{W}_V \mathbf{X},
\]
where \(\mathbf{W}_Q, \mathbf{W}_K, \mathbf{W}_V \in \mathbb{R}^{d \times d_k}\) are parameter matrices, with $d_k$ the dimension of each token after the mapping.

\textit{Computing Attention Matrix} - The attention matrix is computed by using the scaled dot-product similarity between queries and keys and softmax function, i.e.,
\[
\mathbf{A} 
= \text{softmax}\Bigl(\frac{\mathbf{Q} \mathbf{K}^T}{\sqrt{d_k}}\Bigr).
\]

\textit{Multiplying Attention Matrix by Values} - The final self-attention representation is obtained by multiplying the attention matrix with the value vectors:
\[
\mathbf{Y} = \mathbf{A} \mathbf{V}.
\]

\textbf{Multi-Head Attention.} Instead of a single attention computation, Multi-Head Attention \cite{vaswani2017attention} splits the input into $h$ independent attention heads, each with its own set of projections, i.e., $\forall j \in \{1,2,...,h\}$,
\[
\mathbf{Q}_j = \mathbf{W}_{Q_j} \mathbf{X}, \quad
\mathbf{K}_j = \mathbf{W}_{K_j} \mathbf{X}, \quad
\mathbf{V}_j = \mathbf{W}_{V_j} \mathbf{X}.
\]
The attention outputs from all heads are concatenated and projected back:
\[
\mathbf{Y}_{\text{concat}} = [\mathbf{Y}_1; \mathbf{Y}_2; \dots; \mathbf{Y}_h], \quad
\mathbf{Y'} = \mathbf{Y}_{\text{concat}} \mathbf{W}_O,
\]
where $\mathbf{W}_O \in \mathbb{R}^{hd_k \times d_k}$ is a parameter matrix.

\textbf{Residual Connection and Layer Normalization.} In the Transformer architecture, a residual connection \cite{he2016deep} is employed each around each of
the two sub-layers, e.g., Multi-Head Attention and FFN, followed by layer normalization \cite{ba2016layer}. Let \(\mathbf{X}_{\text{in}}\) denote the input to a sub-layer, and let the sub-layer be represented as a function \(f(\cdot)\). The output of the residual block is computed as:

\[
\mathbf{Z} = \text{LayerNorm}(\mathbf{X}_{\text{in}} + f(\mathbf{X}_{\text{in}})).
\]
This design helps stabilize training, improves gradient flow, and enables deeper model architectures.

\textbf{Feed-Forward Network.}
Following multi-head attention, a position-wise feed-forward network (FFN) is applied to each position independently. A common choice is a two-layer fully connected network with a ReLU (or GELU) activation:
\[
\mathbf{FFN}(\mathbf{Z}) = \max(0, \mathbf{Z}\mathbf{W}_1 + \mathbf{b}_1)\,\mathbf{W}_2 + \mathbf{b}_2,
\]
where $\mathbf{W}_1$ and $\mathbf{b}_1$ are the weights and bias of the first linear layer, $\mathbf{W}_2$ and $\mathbf{b}_2$ are the weights and bias of the second linear layer. This step further transforms token representations before passing them to the next layer.

In summary, the Transformer model builds contextualized representations by stacking multiple self-attention layers, each followed by a FFN, residual connections, and layer normalization.

\subsection{Fundamentals of QML}\label{sec2.2}

\subsubsection{Basics}\label{sec2.2.1}

\textbf{Quantum States.} In quantum computing, quantum information is encoded in quantum states defined over a Hilbert space $\mathbb{C}^{2^n}$. A pure $n$-qubit quantum state can be expressed as a unit vector:
$$
|\psi\rangle = \sum_{i=0}^{2^n - 1} \alpha_i |i\rangle,
$$
where $\alpha_i \in \mathbb{C}$ are complex amplitudes satisfying the normalization condition $\sum_i |\alpha_i|^2 = 1$ \cite{Nielsen2002Quantum}.
For a single qubit, the state reduces to:
$$
|\psi\rangle = \alpha |0\rangle + \beta |1\rangle,
$$
with $|\alpha|^2 + |\beta|^2 = 1$, representing a superposition over two basis states.

\textbf{Data Encoding.} To utilize quantum algorithms for processing classical data, the data must first be encoded into quantum states. This process is known as Data Encoding (DE), and it plays a foundational role in quantum machine learning. Two widely adopted encoding strategies are Qubit Encoding (QE) and Amplitude Encoding (AE) \cite{bib34}.

In QE (also called Angle Encoding), classical scalars $x \in \mathbb{R}$ are encoded via parameterized single-qubit rotation gates. For instance, applying a $R_y(x)$ gate to the initial state $|0\rangle$ yields:
$$
|\psi(x)\rangle = \cos\left(\frac{x}{2}\right)|0\rangle + \sin\left(\frac{x}{2}\right)|1\rangle.
$$
Multiple values can be encoded across different qubits by applying independent rotations in parallel.

In AE, a normalized vector $\mathbf{x} = (x_0, x_1, \ldots, x_{2^n-1})$, with $\sum_i |x_i|^2 = 1$, is embedded directly into a quantum state over $n$ qubits:
$$
|\psi(\mathbf{x})\rangle = \sum_{i=0}^{2^n - 1} x_i |i\rangle.
$$
This encoding is highly efficient in space but often requires complex circuit constructions \cite{ashhab2022quantum}.

Other encoding schemes, such as Unary Encoding (UE) and quantum pixel representation (QPIXL) are also used in specific contexts and will be discussed where relevant.

\textbf{Quantum Gates.} Quantum gates are the fundamental operations that evolve quantum states \cite{barenco1995elementary}. They fall into two categories: 1) single-qubit gates, such as rotation gates $R_x, R_y, R_z$, which control the phase and amplitude of individual qubits, and 2) multi-qubit gates, such as the CNOT and CZ gates, which create entanglement between qubits—a key feature enabling quantum parallelism and non-classical correlations. These gates form the building blocks for constructing variational circuits, quantum neural networks, and other hybrid quantum-classical models.

\textbf{Quantun Circuit.} A quantum circuit is a computational module that processes quantum information using a sequence of quantum gates acting on qubits. Mathematically, a quantum circuit applies a unitary transformation \( U \) to an initial quantum state \( |\psi_0\rangle \), producing an output state:  
\begin{equation}  
|\psi_{\text{out}}\rangle = U |\psi_0\rangle.
\end{equation} 
The unitary operator \( U \) is typically decomposed into a series of elementary quantum gates, such as Hadamard, Pauli, and controlled gates, which manipulate the quantum state according to the principles of quantum mechanics \cite{barenco1995elementary}. A general quantum circuit with multiple layers can be expressed as:  
\begin{equation}  
U = U_L U_{L-1} \cdots U_2 U_1,
\end{equation} 
where each \( U_i \) represents a set of quantum gates applied at layer \( i \).  

\textbf{Measurement.} Quantum measurements are described by a collection $\{M_m\}$ of measurement operators. These are operators acting on the state space of the system being measured. The index $m$ refers to the measurement outcomes that may occur in the experiment. If the state of the quantum system is $|\psi\rangle$ immediately before the measurement then the probability that result $m$ occurs is
\begin{equation}
Pr(m) = \langle\psi|M_m^\dagger M_m|\psi\rangle,  \label{eq3-meassure}
\end{equation}
where $M_m^\dagger$ is the conjugate transpose of $M_m$.

\textbf{PQC.} A \textit{Parameterized Quantum Circuit (PQC)} extends the concept of a quantum circuit by incorporating quantum gates with trainable parameters, such as rotation angles in Pauli rotation gates \( R_\theta = e^{-i\theta \sigma/2} \), where $\theta$ controls the rotation angle around a given axis defined by the Pauli matrix $\sigma$. A PQC with $L$ layers can thus be written as:
\begin{equation}
U(\boldsymbol{\theta}) = U_L(\theta_L) \cdots U_2(\theta_2) U_1(\theta_1),
\end{equation}
where each $U_i(\theta_i)$ represents a unitary subcircuit composed of parameterized gates at layer $i$. These parameters effectively define a family of unitary transformations, enabling the circuit to adapt its behavior based on data or task objectives.

PQCs form the backbone of Variational Quantum Algorithms (VQAs) \cite{cerezo2021variational}, which are hybrid quantum-classical frameworks designed for near-term quantum devices. In a typical VQA, the classical data is firstly encoded into a initial quantum state $|\psi_0\rangle$, then a PQC is employed to obtain the final quantum state $|\psi(\boldsymbol{\theta})\rangle = U(\boldsymbol{\theta})|\psi_0\rangle$, and a cost function—usually the expectation value of a Hermitian operator (observable)—is evaluated as:
\begin{equation}
C(\boldsymbol{\theta}) = \langle\psi(\boldsymbol{\theta})| H |\psi(\boldsymbol{\theta})\rangle.
\end{equation}

\begin{figure}[h]
    \centering
    \includegraphics[width=0.48\textwidth]{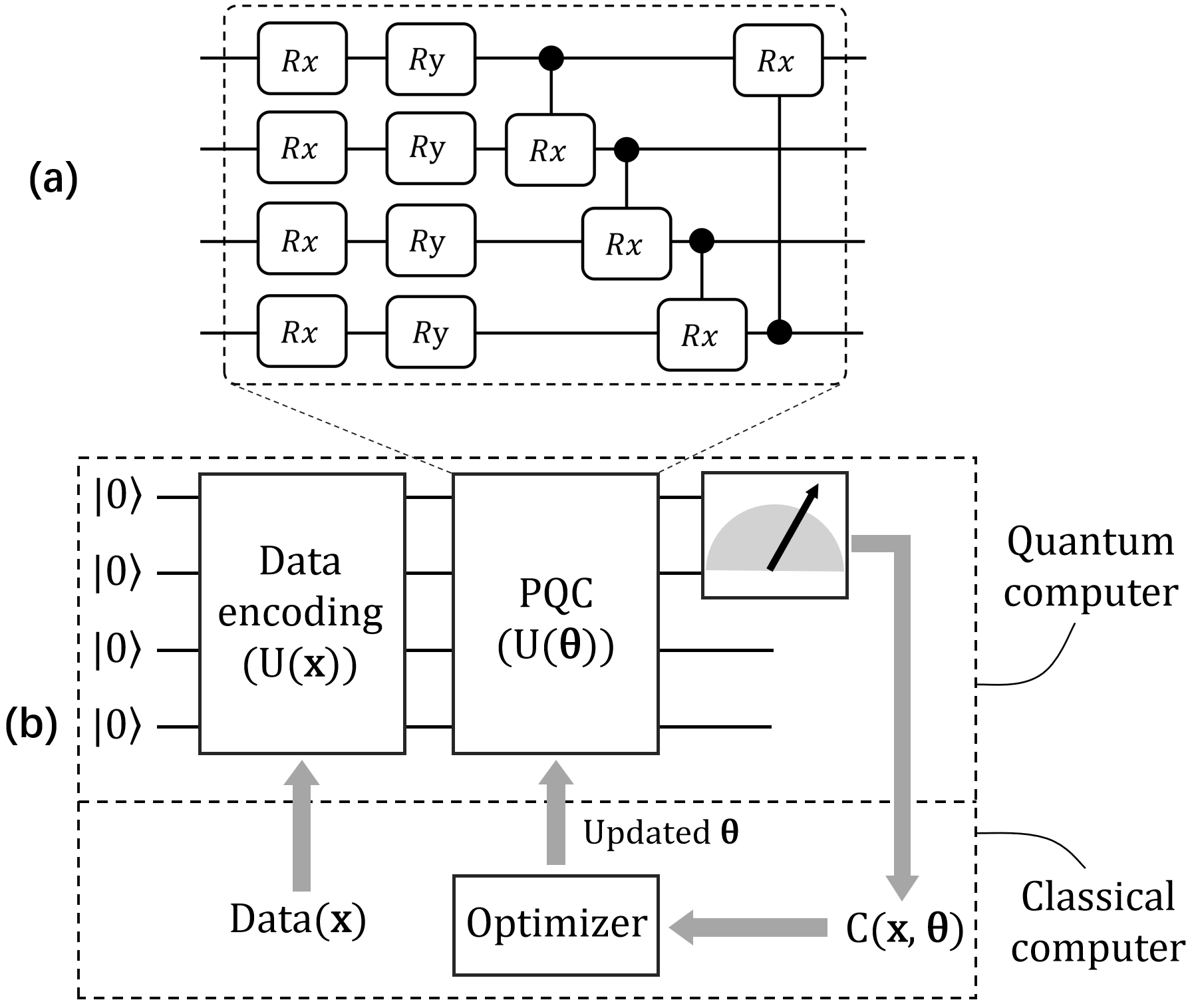}
    \caption{(a) An Example PQC structure. (b) The framework of and VQA.}\label{PQC}
\end{figure}

A classical optimizer is then used to update $\boldsymbol{\theta}$ iteratively to minimize (or maximize) $C(\boldsymbol{\theta})$. This optimization loop combines the expressiveness of quantum circuits with the flexibility of classical learning techniques. Fig. \ref{PQC} illustrates the general workflow of VQA and a typical PQC structure.

\subsubsection{Quantum Tools}\label{sec2.2.2}

\begin{figure}[h]
    \centering
    \includegraphics[width=0.3\textwidth]{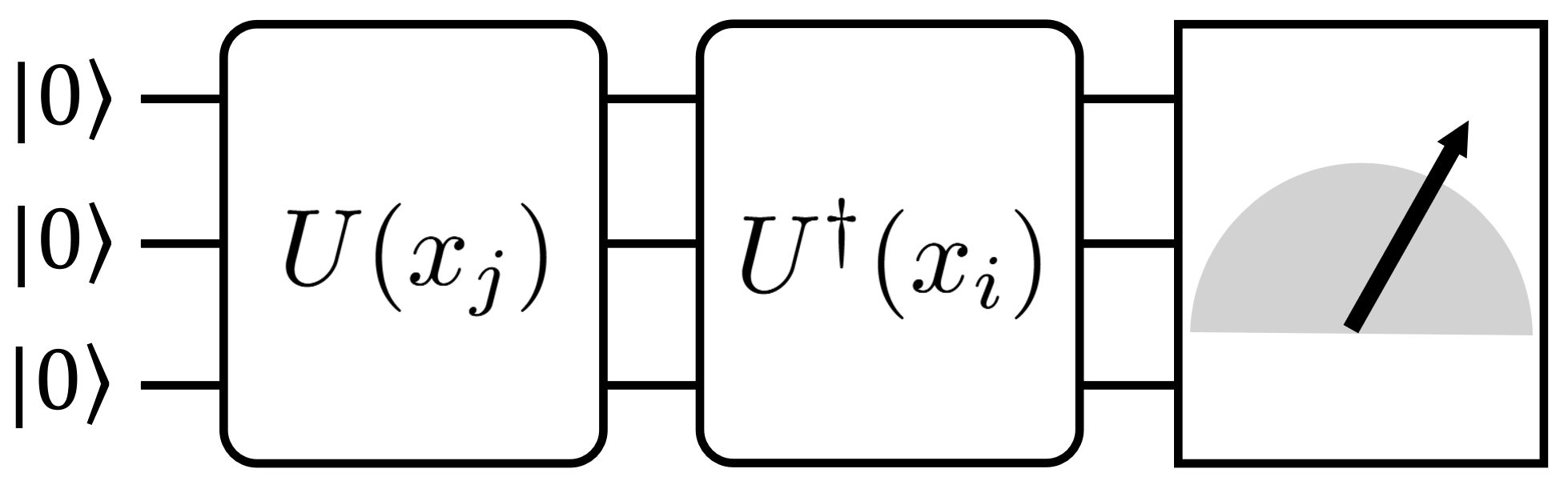}
    \caption{The quantum kernel circuit.}\label{quantum kernel}
\end{figure}

\textbf{Quantum kernel.} Quantum kernel methods leverage quantum computing to enhance classical kernel-based machine learning algorithms, such as support vector machines \cite{bib9}. The core idea is to map classical input data \( x \) into a high-dimensional Hilbert space by using a quantum feature map \( U(x) \), where the inner product between two quantum-encoded data points defines the kernel function:  
\begin{equation}
K(x_i, x_j) = |\langle \psi(x_i) | \psi(x_j) \rangle|^2,
\end{equation}
where $|\psi(x_j) \rangle = U(x_j)|0\rangle^{\otimes n}$, $\langle \psi(x_i)| = \langle0|^{\otimes n}U^{\dagger}(x_i)$.
This quantum kernel measures the similarity between two data points in the quantum feature space, potentially enabling more expressive representations and improved classification performance over classical kernels. The quantum kernel circuit is shown in Fig \ref{quantum kernel}.

\begin{figure}[h]
    \centering
    \includegraphics[width=0.3\textwidth]{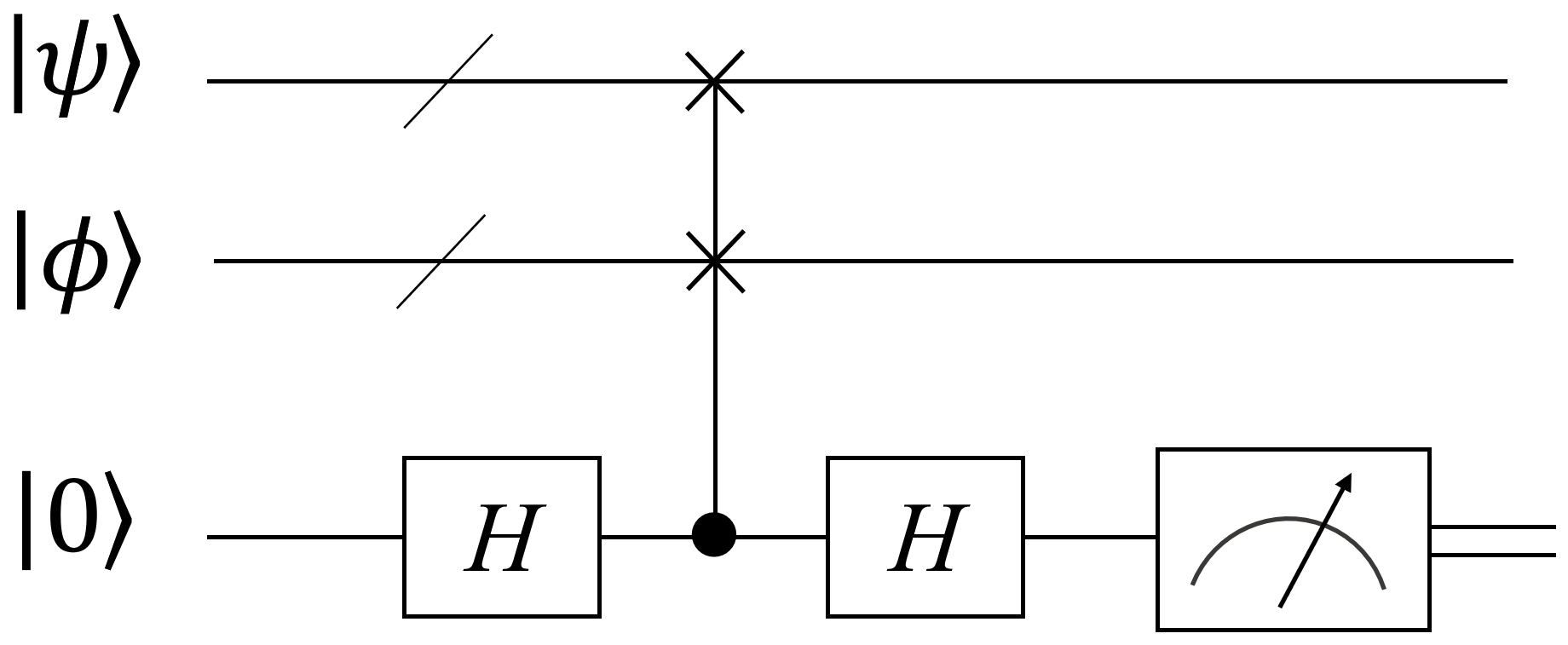}
    \caption{The swap test circuit.}\label{fig-swaptest}
\end{figure}

\textbf{Swap Test.} A swap test is a quantum operation used to determine the similarity between two quantum states \(|\psi\rangle\) and \(|\phi\rangle\) \cite{buhrman2001quantum}. It involves applying a swap operation on the two quantum states, while the operation is controlled by an ancilla qubit. On the ancilla qubit, a Hadamard gate is applied before and after the controlled swap operation. The ancilla qubit is then measured, and the probability of the measurement yielding $0$ is:
\begin{equation}
Pr(0) = \frac{1 + |\langle \psi | \phi \rangle|^2}{2}.
\end{equation}
This probability reflects the degree of overlap between the two states, serving as a measure of similarity. It provides an efficient means of comparing two quantum vectors. The swap test circuit is shown in Fig \ref{fig-swaptest}.

\begin{figure}[h]
    \centering
    \includegraphics[width=0.3\textwidth]{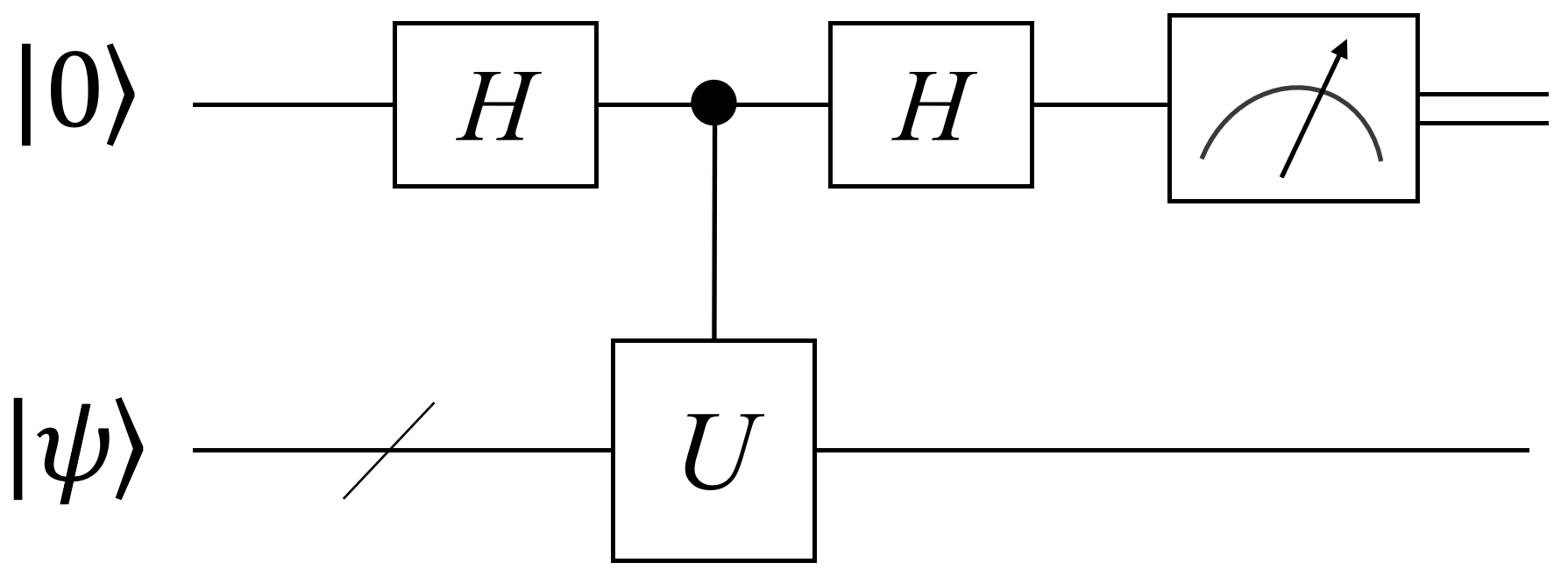}
    \caption{The Hadamard test circuit.}\label{fig-hadamardtest}
\end{figure}

\textbf{Hadamard test.} The Hadamard test is a quantum operation used to estimate the real or imaginary part of the expectation value of a unitary operator \( U \) with respect to a quantum state \( |\psi\rangle \) \cite{montanaro2016quantum}. It utilizes an ancilla qubit to control the application of \( U \) and employs Hadamard gates to create and interfere quantum superpositions. The probability of measuring the ancilla in the \( |0\rangle \) state is: 
\begin{equation}
Pr(0) = \frac{1 + \operatorname{Re} \langle \psi | U | \psi \rangle}{2}.
\end{equation}
Similarly, by modifying the circuit with an additional phase gate, the test can be used to extract the imaginary part of \( \langle \psi | U | \psi \rangle \). This technique is widely used to evaluate inner products, estimate expectation values, and facilitate quantum variational methods. The Hadamard test circuit is shown in Fig \ref{fig-hadamardtest}.

\textbf{Block Encoding.} Block encoding is a quantum technique that embeds a given matrix \( A \) into a larger unitary matrix \( U_A \), thus enabling efficient quantum processing of matrix operations \cite{chakraborty2018power}. It is fundamental in quantum linear algebra and forms the basis for QSVT and quantum algorithms for solving linear systems.

Formally, a unitary matrix \( U_A \) is said to be an (\(\alpha, a, \epsilon\))-block encoding of a matrix \( A \in \mathbb{C}^{2^n \times 2^n} \) if it satisfies the following condition:
\begin{equation}
\|A - \alpha \big( \langle 0^a | \otimes I_n \big) U_A \big( | 0^a \rangle \otimes I_n \big) \| \leq \epsilon, \label{block-encoding}
\end{equation}
where \( \alpha \) is a scaling factor ensuring that \( U_A \) remains unitary, \( a \) represents the number of ancilla qubits required for encoding, \( I_n \) is the identity matrix of appropriate size, and \( \epsilon \) is the allowable error in approximation. Block encoding allows quantum computers to efficiently perform matrix operations, achieving significant speedups in solving complex linear algebra problems over classical methods.

\textbf{Quantum Singular Value Transformation (QSVT).} It is a powerful framework that generalizes various quantum algorithms for linear algebra and matrix computation \cite{gilyen2019quantum}. It enables the manipulation of the singular values of a block-encoded matrix using quantum circuits, providing exponential speedups for such problems as solving linear systems, matrix exponentiation, and principal component analysis over classical methods.

Given a block-encoded matrix \( U_A \) of a target matrix \( A \), QSVT applies a carefully designed sequence of quantum operations to transform the singular values \( \sigma_i \) of \( A \). Formally, if \( A \) is (\(\alpha, m\))-block-encoded in \( U_A \), QSVT constructs a polynomial transformation \( P(A) \) such that:
\begin{equation}
P(A) = V_A P(D_A) V_A^\dagger,
\end{equation}
where \( D_A \) is a diagonal matrix containing the singular values \( \sigma_i \) of \( A \), and \( V_A \) is a unitary transformation that diagonalizes \( A \). The polynomial \( P \) is designed through a sequence of phase rotations and controlled operations, allowing for the controlled amplification, filtering, or inversion of singular values.

A key advantage of QSVT is its ability to approximate functions of a matrix \( A \) with minimal overhead, making it a fundamental technique in quantum algorithms for machine learning, optimization, and scientific computing.

\section{Technical Approaches} \label{sec3}

\subsection{PQC-based Quantum Transformers} \label{sec3.1}

In the NISQ era, quantum hardware is limited by high noise levels and a restricted number of qubits. Consequently, most research efforts focus on exploring the feasibility of quantum Transformers within the framework of PQCs. By replacing neural network weight layers with PQCs and integrating specific quantum algorithms, these models aim to reduce classical computational complexity of Transformer while potentially demonstrate quantum advantages. This section analyzes the architecture of these models, evaluates their quantum resource requirements and performance, and lays the foundation for further optimization and extension of quantum Transformers. 

\begin{table*}[h]
\centering
\caption{The classification of PQC-based Quantum Transformers. In the ‘Model Schematic’ column, the purple elements represent quantum operations or quantum state data, while orange elements correspond to classical operations or classical data. Green nodes indicate steps that are implemented differently depending on the method—some use classical processing, while others employ quantum solutions.}
\label{classification of PQC-based models}
\begin{tabular}{C{2.3cm}C{6.2cm}C{3.6cm}C{2.2cm}C{1.5cm}}
\toprule%
\textbf{Category} & \textbf{Model Schematic} & \textbf{Classification principles} & \textbf{Attention mechanism} & \textbf{Ref.} \\ \midrule
QKV-only Quantum mapping & \includegraphics[width=5cm]{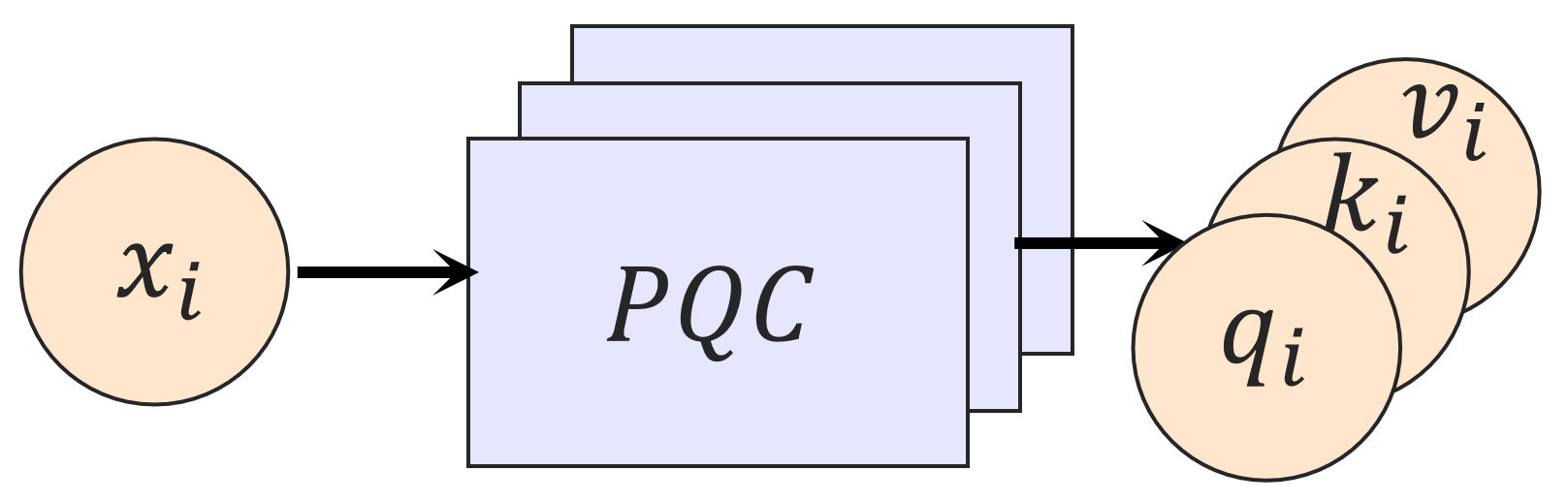} & Only using PQC to map the input \( \{x_i\}_{i=1}^N \) to \( \{q_i\}_{i=1}^N \), \( \{k_i\}_{i=1}^N \), \( \{v_i\}_{i=1}^N \), leaving the attention computation as a classical step. &  $\text{softmax}\left(\frac{QK^\top}{\sqrt{d_k}}\right)$ (classical) & \cite{li2024quantum,zhang2024light-weight,wei2023povm,bib64,cara2024quantum,unlu2024hybrid,chakraborty2025integrating,dutta2024aq,dutta2024qadqn,he2024training} \\ \midrule
Quantum Pairwise Attention (inner-product similarity) & \includegraphics[width=6.7cm]{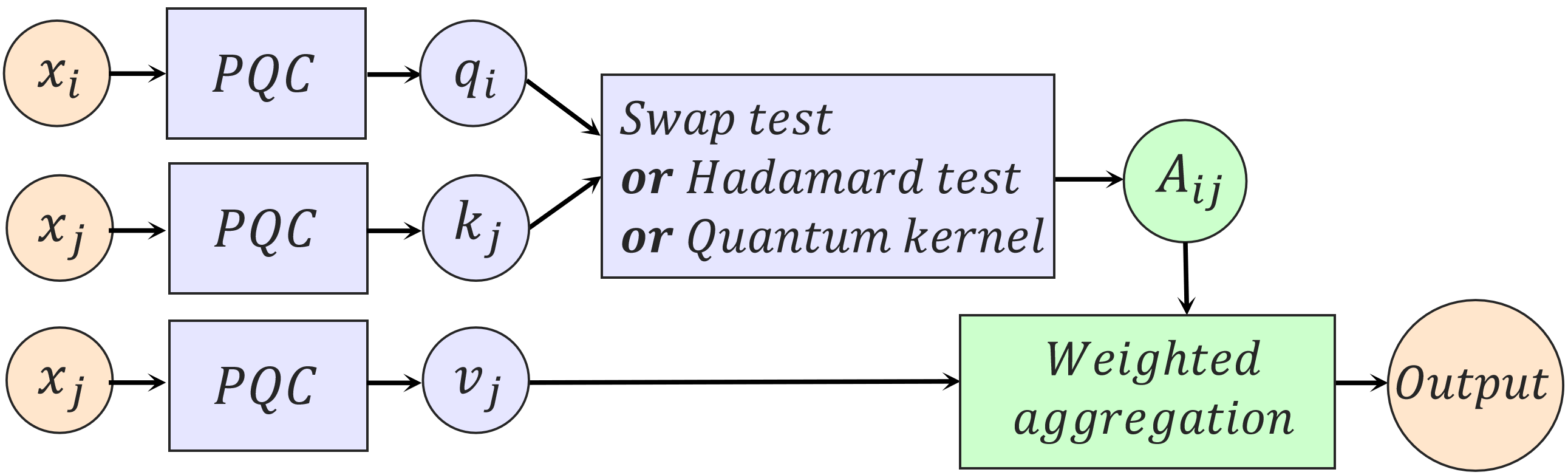} & Pairwise similarity computation between tokens with inner product similarity. & $\{|\langle q_i | k_j \rangle|^2\}_{i,j=1}^N$ or $\{Re\langle q_i | k_j \rangle\}_{i,j=1}^N$ & \cite{chen2025quantum,zhang2025hqvit,smaldone2025hybrid,zhao2024QKSAN,kamata2025molecular} \\ \midrule
Quantum Pairwise Attention (generalized similarity) & \includegraphics[width=6.7cm]{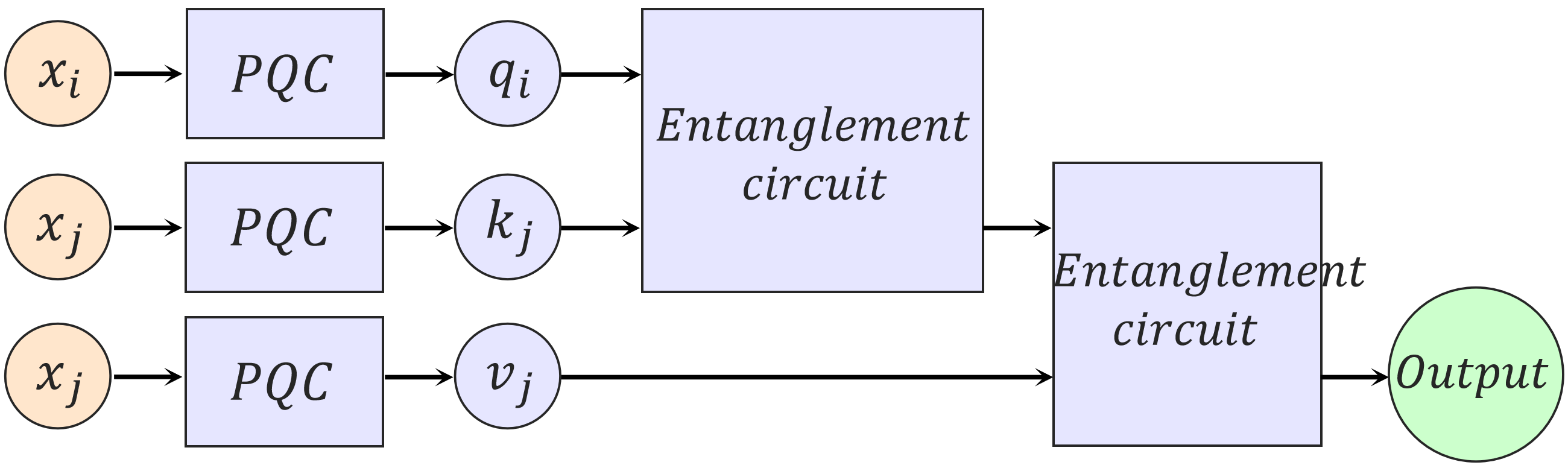} & Pairwise similarity computation between tokens with generalized similarity metric. & $\{f(q_i, k_j)\}_{i,j=1}^N$ & \cite{shi2024QSAN,zheng2023Design,shi2023natural,chen2025quantum-c} \\ \midrule
Quantum Holistic Attention & \includegraphics[width=6.7cm]{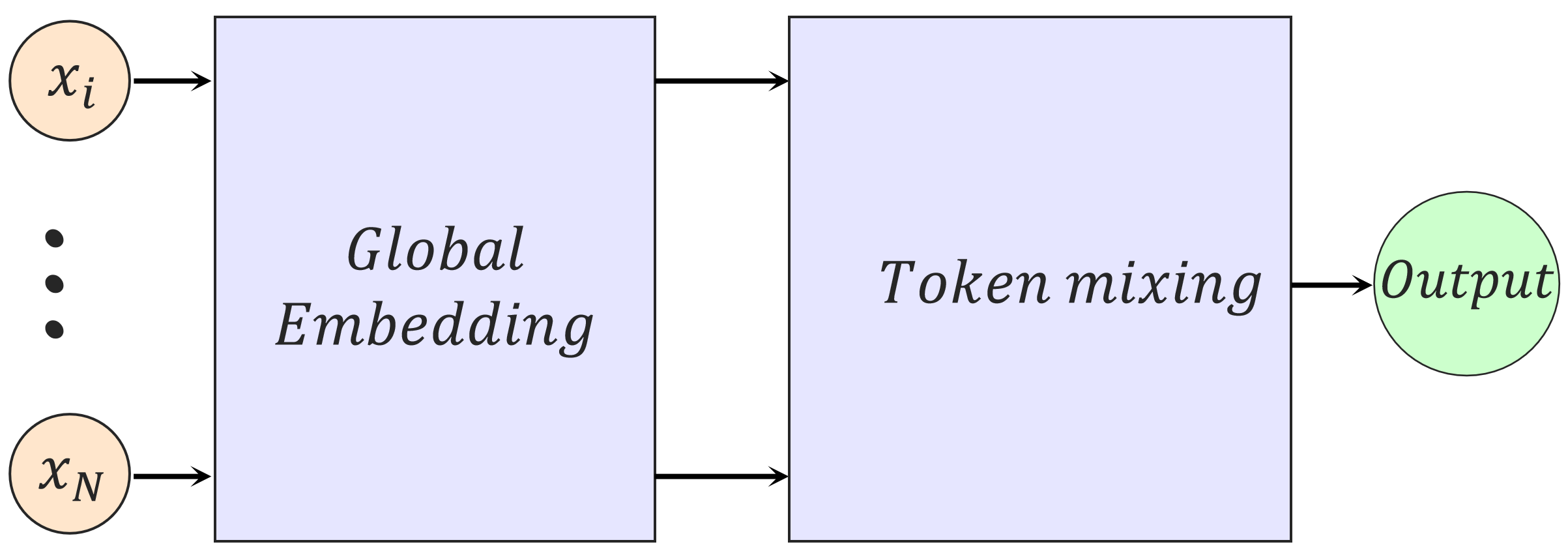} & Captures global token dependencies through holistic quantum transformations, without explicitly computing pairwise similarities. & $\mathcal{U}_\text{mix}(\{x_i\}^N_{i=1})$ & \cite{kerenidis2024quantum,tesi2024quantum,evans2024learning} \\ \midrule
Quantum-Assisted Optimization & \includegraphics[width=6.5cm]{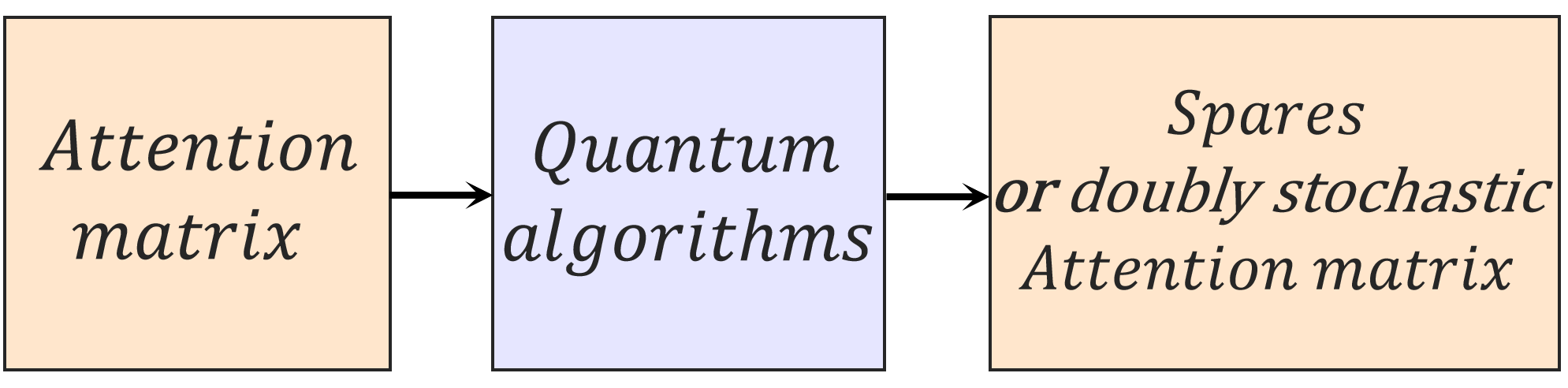} & Using quantum algorithms to optimize the attention matrix. & Sparsification / Doubly stochastic process & \cite{gao2023fast, born2025quantum} \\ \bottomrule
\end{tabular}
\end{table*}

We examine these PQC-based model architectures through the lens of quantumization strategies, dividing them into four categories as shown in Table \ref{classification of PQC-based models}: QKV-only Quantum mapping (quantumizing solely \(Q/K/V\) mapping step), Quantum Pairwise Attention methods (preserving the pairwise token similarities form), Quantum Holistic Attention (implicit global token mixing without $Q/K/V$), and Quantum-Assisted Optimization (the quantum algorithms serve as an assisted acceleration and do not replace the classical components). This taxonomy elucidates variations among these models in a quantumization degree (ranging from shallow quantum to fully quantum), optimization aims (replacement vs. optimization), and implementation strategies (classical retention vs. quantum reinvention), thereby offering a comprehensive technical roadmap for current PQC-based Quantum Transformers.

\begin{figure*}[t]
    \centering
    \includegraphics[width=0.99\textwidth]{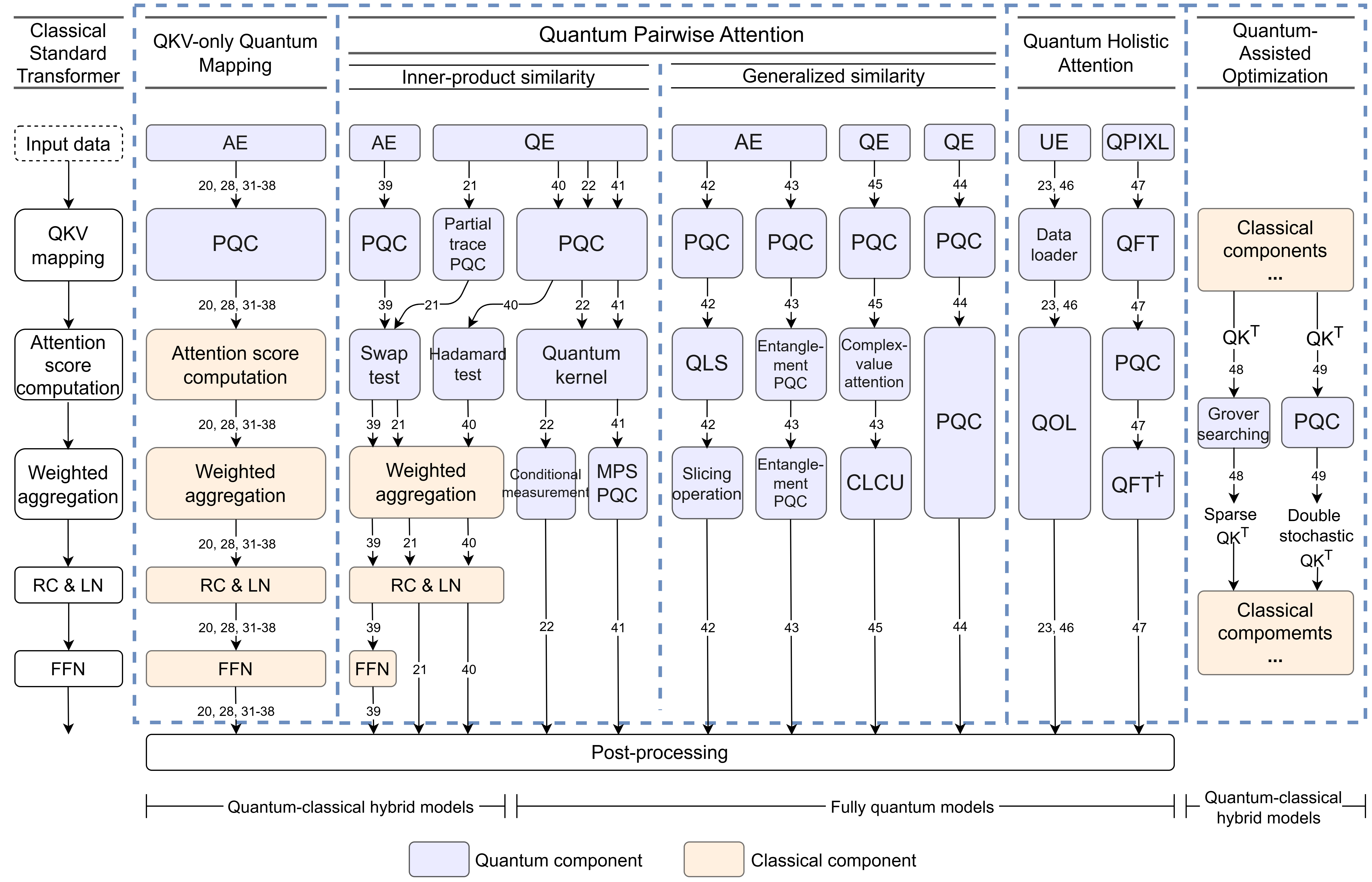}
    \caption{Technical roadmap of PQC-based quantum Transformer architectures under four quantumization strategies. Columns correspond to the four-category classification, while rows align with the standard Transformer pipeline—from input encoding to output classification. The figure highlights where quantum components intervene, the specific quantum tools applied, and the extent of classical-quantum hybridization. The quantum-assisted optimization (rightmost column) functions as an auxiliary module, rather than an in-line replacement. Colors distinguish quantum (purple) and classical (orange) modules. Reference numbers indicate exemplar studies implementing each approach.} \label{roadmap}
\end{figure*}

We further compile Fig. ~\ref{roadmap}, which presents a more fine-grained technical roadmap aligning all these models with the standard Transformer architecture. Each column corresponds to one of the four categories of PQC-based quantum transformers, while each row aligns with a functional stage—from data encoding to FFN. This layout highlights where quantum operations are inserted, how they interact with classical components, and what quantum tools (e.g., PQC, swap test, QFT, Grover search) are used. Reference numbers point to specific models, bridging abstract strategies with concrete implementations and offering a practical view of the quantumization landscape.

\subsubsection{QKV-only Quantum mapping}\label{sec3.1.1}

QKV-only Quantum mapping methods refers to quantumizing only the \(Q, K, V\) mapping step in Transformers, replacing classical linear mapping matrices (e.g., $W_Q$, $W_K$ and $W_V$) with PQCs, while retaining core self-attention computations and subsequent steps as classical implementations, as shown in Table. \ref{classification of PQC-based models}. This approach generates enhanced feature representations by using quantum Hilbert space, aiming to enrich \(Q/K/V\)'s expressivity while maintaining compatibility with classical Transformers. A summary table (Table \ref{Summary for QKV-only}) is provided to highlight the key features of all works in this category, including insights, contributions, and limits.

\textbf{A. Representative work}

\begin{table*}[htbp]
\centering
\renewcommand{\arraystretch}{1.2} % Reduces vertical spacing (adjust value as needed)
\caption{Summary for the key features of the QKV-only Quantum Mapping methods} \label{Summary for QKV-only}
\begin{tabular}{|>{\raggedright\arraybackslash}p{1.5cm}|m{6cm}|m{4.5cm}|m{4cm}|}
\hline
\multicolumn{1}{|c|}{\textbf{Ref.}} & \multicolumn{1}{c|}{\textbf{Insight}} & \multicolumn{1}{c|}{\textbf{Contributions}} & \multicolumn{1}{c|}{\textbf{Limits}} \\ \hline
\cite{li2024quantum} & $\bullet$ Using PQC for QKV quantum mapping & $\bullet$ Enhancing linear mapping expressiveness through quantum space & $\bullet$ QK as scalar, limited expressivity \\ 
      & $\bullet$ Using GPQSA to overcome the limitations of unitary transformations & $\bullet$ Highly flexible and adaptable model & $\bullet$ Remaining $O(N^2)$ Classical computational overhead \\ \hline
\cite{zhang2024light-weight} & $\bullet$ Using APDM to decouple the measurement results into different classical feature spaces  & $\bullet$ More lightweight model & $\bullet$ Remaining $O(N^2)$ Classical computational overhead \\ 
      & $\bullet$ Improving the structure of PQC & $\bullet$ Stronger QK representation &  \\ \hline
\cite{wei2023povm} & $\bullet$ Extracting QKV information with POVM & $\bullet$ Stronger QK representation & $\bullet$ Scalability needs further validation \\ \hline
\cite{bib64} & $\bullet$ Embedding Quantum Transformer into KNN clustering & $\bullet$ Better performance than classical baseline & —— \\ \hline
\cite{cara2024quantum,unlu2024hybrid,dutta2024aq,dutta2024qadqn,chakraborty2025integrating} & $\bullet$ Extending QSANN \cite{li2024quantum} to multi-head attention or Embedding it into a full Transformer architecture & $\bullet$ Real-world application, comparable to classical ViT & $\bullet$ Underperforms classical models in certain contexts, with restricted generalization capability \\ \hline
\cite{he2024training} & $\bullet$ Executing QSANN \cite{li2024quantum} on a real quantum computer & $\bullet$ Validation of QSANN's NISQ feasibility & —— \\ \hline
\end{tabular}
\end{table*}

Li et al. \cite{li2024quantum} are the first to propose a Quantum Self-Attention Neural Network (QSANN) based on this concept. QSANN firstly encodes the classical data into quantum state,
\begin{equation}
|\psi(x)\rangle = U(x) |0\rangle^{\otimes n},
\end{equation}
then performs PQCs on the initial quantum state separately,
\begin{equation}
|\psi_i\rangle = U_i(\theta) |\psi(x)\rangle, i \in \{Q, K, V\}.
\end{equation}
Then, measurements are performed to obtain three sets of expectations, which are used as the mapped $QKV$. In this process, the three quantum systems of $QKV$ are independent of each other. To address the challenge of correlating distant quantum states, Li et al. \cite{li2024quantum} introduce the Gaussian Projected Quantum Self-Attention (GPQSA) mechanism. This mechanism calculates self-attention coefficients through a gaussian projected kernel, rather than relying on traditional inner products. Consequently, the combination of quantum generation operations with the classical self-attention mechanism forms a quantum-classical hybrid self-attention layer to extract the feature representations of the input data. These feature representations are then averaged and used for classification tasks. Experimental results demonstrate that QSANN achieves higher classification accuracy than those of the traditional models in binary classification tasks on small-scale text datasets, highlighting the potential of quantum-enhanced self-attention in practical applications.

Due to its simple structure and ease of integration with classical models, QSANN has attracted considerable attention and inspired a series of follow-up studies \cite{zhang2024light-weight, wei2023povm, unlu2024hybrid, cara2024quantum, bib64, he2024training, chakraborty2025integrating, dutta2024qadqn}. These studies further refine measurement methods, improve PQC structures, or apply QSANN to specific practical problems.

\textbf{B. Measurement Improvement}

Building upon QSANN\cite{li2024quantum}, Zhang et al. \cite{zhang2024light-weight} propose an improved quantum Transformer model that introduces Amplitude-Phase Decomposed Measurements (APDM) and more expressive PQCs. In this approach, each PQC is measured under both Pauli-X and Pauli-Z bases, and the measurement results are interpreted as different classical features. This enables the mapping of Q, K, and V with just two PQCs, reducing the number of learnable parameters by one-third while improving information extraction efficiency. In parallel, Wei et al. \cite{wei2023povm} employe Positive Operator-Valued Measurement (POVM) \cite{Nielsen2002Quantum} as a readout mechanism. By leveraging informationally complete (IC) tetrahedral POVM operators on each qubit to map quantum states of Q, K, and V into classical space, their approach provides a richer and more isotropic sampling of the Bloch sphere, potentially leading to more expressive classical representations from quantum features. Experimental evaluations in both works report slightly improved performance over the QSANN \cite{li2024quantum}, suggesting that advanced measurement strategies are important to unlocking the representational power of quantum Transformers.

\textbf{C. Practical Application}

Since the $QKV$ components are converted into classical data for subsequent processing, the QKV-only Quantum mapping approach can be naturally extended to multi-head attention, thereby enhancing model performance while maintaining compatibility with standard Transformer architectures.

This design flexibility has enabled a variety of downstream applications. For instance, \cite{unlu2024hybrid} and \cite{cara2024quantum} adopt QKV-only quantum mapping models for high-energy physics image classification, demonstrating competitive performance in complex visual tasks. In the domain of time-series forecasting, similar frameworks have been employed by \cite{chakraborty2025integrating} and \cite{dutta2024qadqn}, showing the feasibility of quantum-enhanced attention for temporal data modeling. In the context of physics-informed modeling, \cite{dutta2024aq} employs the QKV-only quantum mapping approach to assist in numerically solving the Navier–Stokes equations in fluid dynamics.

Beyond regression and prediction tasks, quantum self-attention has also been explored for unsupervised learning. Nguyen et al. \cite{bib64} propose a quantum Transformer model, QClusformer, which refines image clustering by computing inter-feature correlations via a quantum self-attention layer. This mechanism helps identify hard samples and noise within coarse-grained clusters initially formed by a classical $k$-nearest neighbor algorithm.

Together, these practical efforts across multiple domains demonstrate the versatility and effectiveness of PQC-based linear quantum mappings in processing high-dimensional data.

\textbf{D. Experiments on a Real Quantum Computer}

He et al. (2024) \cite{he2024training} pioneer the execution of a quantum Transformer model on real quantum hardware, with an overall architecture resembling QSANN but featuring slight improvements in the structure of PQCs to ensure that single-qubit measurements adequately capture complex dependencies. The experiments utilize the "Wukong" 72-qubit superconducting quantum computer, employing 28 of its qubits. Wukong, developed by Origin Quantum in collaboration with the Chinese Academy of Sciences, is one of China's most advanced superconducting quantum processors \cite{chinadaily2023wukong}. By introducing parallel strategies at the attention, and batch levels, this work theoretically accelerates training speed by a factor of \(3 \times bs \times n\) compared to non-parallel strategies. 
The method’s effectiveness was validated on the MC and RP datasets \cite{lorenz2023qnlp}, with results showing that performance on the real quantum chip (MC accuracy 100\% and RP accuracy 83.87\%) slightly outperformed the simulator (MC 100\%, RP 80.66\%), while significantly reducing forward and backward propagation time. This work not only demonstrates the feasibility of quantum self-attention models on NISQ devices but also provides valuable insights for the practical deployment of QML algorithms in NLP tasks through parallel optimization and noise-adaptive design. Future research could further explore its scalability and applicability to more complex tasks.

Although the QKV-only Quantum mapping approach is favored for its simple and flexible structure, it have some limitations. First, it does not address the computational bottleneck in the classical transformer model—the computation of the attention matrix. Second, when stacking multiple self-attention layers, data must be frequently converted between the quantum and classical state, which increases the overhead of data encoding and measurements. Therefore, more studies seek deeper quantum applications to better highlight the advantages of quantum computing.

\subsubsection{Quantum Pairwise Attention}\label{sec3.1.2}

This section focuses on quantum pairwise attention methods, which preserve the form of computing pairwise similarities between two tokens. It is worth noting that this subclass of work represents a deeper level of quantumization compared to the first subclass, as $Q, K$, and $V$ are also generated by PQCs. The key difference is that, in this routine, $Q$, $K$, and $V$ remain as quantum states rather than being measured before attention score computation. Such methods can be further categorized into quantum pairwise inner product similarity methods and generalized ones. A summary table (Table \ref{Summary for Quantum Pairwise Attention}) is provided to highlight the key features of all works in this category, including insights, contributions, and limits.

\begin{table*}[htbp]
\centering
\renewcommand{\arraystretch}{1.2} % Reduces vertical spacing (adjust value as needed)
\caption{Summary for the key features of the Quantum Pairwise Attention method} \label{Summary for Quantum Pairwise Attention}
\begin{tabular}{|m{1cm}|m{6cm}|m{5.5cm}|m{3.5cm}|}
\hline
\multicolumn{1}{|c|}{\textbf{Ref.}} & \multicolumn{1}{c|}{\textbf{Insight}} & \multicolumn{1}{c|}{\textbf{Contributions}} & \multicolumn{1}{c|}{\textbf{Limits}} \\ \hline
\cite{chen2025quantum} & $\bullet$ Using swap test to obtain inner product & $\bullet$ Lower inner product complexity & $\bullet$ Requires $n^2$ sub-circuit runs \\ 
      & $\bullet$ Obtain mixed-state QK representation via partial trace & $\bullet$ Enhanced QK representation (Overcoming unitary transform constraints) &  \\ \hline
\cite{zhang2025hqvit} & $\bullet$ Inputting whole image with AE encoding, encoding both pixel and positional information & $\bullet$ Qubit resource efficiency & $\bullet$ More samplings needed for conditional measurement \\ 
      & Using swap test to compute inner product & $\bullet$ Lower inner product complexity & Requires $n^2$ sub-circuit runs \\ \hline
\cite{smaldone2025hybrid} & $\bullet$ Using Hadamard test to obtain inner product & $\bullet$ Lower inner product complexity & $\bullet$ Requires $n^2$ sub-circuit runs \\
& $\bullet$ Applying Quantum Transformer for generative tasks & $\bullet$ Better generation results than classical baseline & \\ \hline
\cite{zhao2024QKSAN} & $\bullet$ Using Quantum Kernel to obtain pairwise similarity of tokens & $\bullet$ QK shares one register, reducing qubit resource needs & $\bullet$ Requires $n^2$ sub-circuit runs \\
& $\bullet$ Applying quantum state similarity directly to V register via conditional measurement & $\bullet$ Higher-dimensional similarity representation & $\bullet$ More samplings needed for conditional measurement \\ \hline
\cite{kamata2025molecular} & $\bullet$ Using Quantum Kernel to obtain pairwise similarity of tokens & $\bullet$ QK shares one register, reducing qubit resource needs & $\bullet$ Requiring considerable circuit depth \\
& $\bullet$ Getting weighted V by cascading circuits, without classical aggregation & $\bullet$ Reducing measurement count &  \\
& $\bullet$ Embedding Quantum Transformer into a VQE & $\bullet$ Better performance than classical baseline & \\ \hline
\cite{shi2024QSAN} & $\bullet$ Proposing fully quantum similarity representation, i.e., QLS & $\bullet$ Fully quantum, reduced quantum-classical interface conversion & $\bullet$ Significant increase in quantum circuit width and depth \\ \hline
\cite{zheng2023Design} & $\bullet$ Use modular PQC design and entanglement to obtain token similarity and weighted aggregation & $\bullet$ Fully quantum, reduced quantum-classical interface conversion & $\bullet$ Extremely deep quantum circuits required \\ \hline
\cite{shi2023natural} & $\bullet$ Input paired $q_i$, $k_j$, and $v_j$ to quantum circuit, yielding weighted $v_i$ & $\bullet$ Fully quantum and simple architecture & $\bullet$ Repetitive n-time input per token leads to extreme qubit resource consumption \\ \hline
\cite{chen2025quantum-c} & $\bullet$ Propose the complex-valued attention mechanism that leverages both the amplitude and phase information of quantum states & $\bullet$ Enhancing the similarity representation & $\bullet$ Complex circuit structure \\
& $\bullet$ Using CLCU to achieve weight aggregation with quantum manner & $\bullet$ Multi-head quantum attention mechanisms enhance model capabilities & \\ \hline
\end{tabular}
\end{table*}

\textbf{A. Quantum pairwise inner-product similarity}

They replace the classical computation of \(\langle q_i, k_j \rangle\) with quantum methods that can directly obtain the inner-product similarity between two quantum states, e.g., \textbf{Swap Test, Hadamard Test, and Quantum Kernel}. In this section, we introduce the resulting models based on these quantum algorithms.

Chen et al. \cite{chen2025quantum} propose the Quantum Mixed-State Self-Attention Network (QMSAN) by using the swap test to directly compute the similarities of token pairs. This method firstly uses PQCs to generate $|\psi_q\rangle$, $|\psi_k\rangle$ and $|\psi_v\rangle$, then performs partial trace operations on $|\psi_q\rangle$ and $|\psi_k\rangle$ to obtain $|\hat{\psi}_q\rangle$ and $|\hat{\psi}_k\rangle$ in mixed quantum states. Then, the swap tests are performed on this mixed quantum state pairs to obtain the attention matrix. Since the outcomes of swap tests are classical, the subsequent steps are converted to classical means. The use of feature representations based on mixed quantum states allows the evolution from input data to the $Q$, $K$, and $V$ representations to extend beyond purely unitary transformations, enabling more general non-unitary mappings. The feature representation based on quantum mixed states allows the evolution of input data in PQCs to overcome the limitations of unitary transformations, resulting in $Q$ and $K$ with richer features. On small-scale NLP datasets, the model demonstrates superior classification accuracy compared to existing quantum Transformer approaches.

Building on this idea, Zhang et al. \cite{zhang2025hqvit} introduce the Hybrid Quantum Vision Transformer (HQViT), which also employs the swap test for attention computation, but focuses specifically on image data. Unlike patch-based encoding used in classical ViT models, HQViT processes the entire image as a whole via amplitude encoding. This results in a natural partitioning of the quantum system into two subsystems: one encoding the token content (subsystem 1), and the other encoding token index information (subsystem 2). Two such quantum systems are prepared to represent the Q and K matrices, and the swap test operates on subsystem 1. A systematic traversal of conditional measurements is then performed on subsystem 2, allowing for full construction of the attention matrix across all tokens. This method preserves global spatial information and avoids information loss from patching. This method naturally embeds positional information of the image, enhancing the model's contextual representation capability without increasing quantum resource consumption. By achieving a favorable balance between classical and quantum computational resource consumption, the model delivers classification performance on small-scale CV datasets that rivals or surpasses other quantum Transformer models.

In contrast to the swap test, Smaldone et al. \cite{smaldone2025hybrid} explore the use of the Hadamard test for inner product estimation in their hybrid quantum-classical Transformer designed for molecular generation tasks. Building on the conventional Hadamard test, this approach incorporates controlled inversion and conditional reset operations to efficiently embed \(\lvert q_i \rangle\) and \(\lvert k_j \rangle\) and calculate their inner product. The process begins with a primary register preparing \(\lvert q_i \rangle\), followed by a conditional reset of the register under the control of an auxiliary qubit to embed \(\lvert k_j \rangle\), with similarity extracted via a final Hadamard gate and measurement. The attention coefficient is defined as the real part of the inner product, i.e., $\text{Re} \langle q_i \lvert k_j \rangle$. The resulting classical attention matrix is combined with the value matrix \(\mathbf{V}\) to produce the output. On the QM9 dataset, this method generates molecules with target properties, achieving performance comparable to classical Transformers, thus demonstrating its potential on NISQ devices.

Zhao et al. \cite{zhao2024QKSAN} propose the Quantum Kernel Self-Attention Mechanism (QKSAM), which computes the attention coefficients by encoding both \(\langle q_i|\) and \(|k_j\rangle\) into a single quantum register and directly obtains their overlap by measurements. Compared to the swap test and Hadamard test-based approaches \cite{chen2025quantum, zhang2025hqvit, smaldone2025hybrid}, the quantum kernel-based approach reduces qubit resource requirements by one-third when computing similarity, as the evolution of $Q$ and $K$ shares a single quantum system (while the evolution of $V$ still requires an independent one). The attention coefficients are derived through the qubit-wise conditional measurements, which are called Quantum Kernel Self-Attention Score (QKSAS), and then QKSAS are associated to the $V$ register by control operations to generate weighted values. This approach yields a new interpretation of attention coefficients, presenting them as probability vectors rather than scalars, thereby expanding the representation space of similarity coefficients and enhancing the model's expressive power.

Kamata et al. propose the Molecular Quantum Transformer (MQT) \cite{kamata2025molecular}, a variational quantum eigensolver (VQE) \cite{peruzzo2013variational}-based method that replaces the ansatz with a Quantum Transformer architecture. Instead of using a manually designed parametrized circuit structure, this work employs a quantum transformer network to represent the variational wavefunction, enabling the model to learn the ground-state electron configuration that minimizes the energy with respect to a given molecular Hamiltonian. The core component of MQT, i.e., the quantum self-attention layer, utilizes a quantum kernel method to capture the correlations between query \( q_i \) and key \( k_j \), enabling a more effective representation of the complex interactions and correlations between electrons and nuclei. Subsequently, a Matrix Product State (MPS) module compresses the attention coefficients into a single qubit, which is then assigned to the value register \( v_j \). MQT transforms molecular configuration features (such as nuclear coordinates and electron-nucleus distances) into quantum states via angle embedding. These are processed through a quantum Transformer module to generate enhanced molecular structure representations, optimizing the variational estimation of ground-state energies. In terms of performance, MQT demonstrates superior accuracy compared to the classical Transformer in numerical experiments. For instance, on \( \mathrm{LiH} \), \( \mathrm{BeH}_2 \), and \( \mathrm{H}_4 \), MQT reduces the average estimation error by 44\%, 18\%, and 74\%, respectively, while exhibiting robust generalization capabilities.

The swap test, Hadamard test and quantum kernel algorithms can be used to lower the complexity of inner-product computation from $O(d)$ to $O(1)$ (ignoring the sampling count). However, to compute the entire attention score matrix, the circuit still needs to be executed \(N^2\) times. Therefore, the algorithm complexity is \(O(N^2 f(d))\), which still scales quadratically with the sequence length. Here, \(f(d)\) represents the encoding complexity of the circuit, depending on the encoding method. For qubit encoding, \(f(d) = d\), while for amplitude encoding, \(f(d) = \log d\).

\textbf{B. Quantum pairwise generalized similarity}

Instead of using inner-product to be the similarity metric, some studies adopt generalized similarity metric, thus leveraging the unique properties of quantum circuits to implement the pairwise attention mechanism.

Zhao et al. \cite{shi2024QSAN} develop the Quantum Self-Attention Network (QSAN) that introduces Quantum Logical Similarity (QLS). QLS is a new metric that replaces inner-product similarity. QSAN utilizes quantum gates (such as Toffoli and CNOT gates) to perform logical operations and compute the similarity between Query and Key. Compared to classical methods, QLS avoids numerical computations and intermediate measurements, allowing the model to continuously operate on a quantum computer and obtain a similarity representation with quantum characteristics—Quantum Bit Self-Attention Score Matrix (QBSASM). QBSASM represents attention scores in the form of quantum states (tensors), which have a higher dimensionality compared to classical scalar representations, enabling it to capture richer information in Hilbert space. The association between QBSASM and \( V \) is achieved through a slicing operation. Specifically, for each query \( q_i \), the slicing operation extracts the QLS elements, i.e., (\( \langle k_j | q_i \rangle \)), from each row of QBSASM as control bits, then multi-controlled Toffoli gates are then used to apply weighted control over \( v_j \). The model is validated on the MNIST and CIFAR-10 datasets, and experimental results demonstrated its faster convergence and higher classification accuracy. However, this method requires a large number of auxiliary qubits to store intermediate results for the AND and modulo-2 addition of Q-K pairs, causing the model width and depth complexity to grow quadratically with the sequence length \( N \). This results in high quantum resources consumption.

Another quantum pairwise generalized attention model is proposed in \cite{zheng2023Design}. It comprises three main steps: 1) the input data is encoded into quantum states with amplitude encoding, generating \( Q|x\rangle \), \( K|x\rangle \), and \( V|x\rangle \); the quantum self-attention layer utilizes strongly entangled quantum circuit block (composed of parameterized rotation gates and CNOT gates) to capture the similarity between each \( q_i \) and \( k_j \) pair; 2) a universal 2-qubit gate block is applied to distribute this similarity to the corresponding \( v_j \), producing a weighted value; and 3) all the weighted Values are further processed by a quantum fully connected layer (incorporating Hadamard gates, CNOT gates, and rotation gates) for classification, with the entire process operating directly on quantum states without auxiliary qubits. The parameters are optimized by a network optimization module that leverages quantum stochastic gradient descent (QSGD) and the parameter-shift rule. The classification performance of this method outperforms DisCoCat \cite{lorenz2023qnlp} and QSANN \cite{li2024quantum}. Compared to QSAN \cite{shi2024QSAN}, although this method \cite{zheng2023Design} does not require \( \mathcal{O}(N^2) \) auxiliary qubits, the pairwise entanglement between \( q_i \) and \( k_j \) occurs along the circuit depth. As a result, the overall circuit depth remains \( \mathcal{O}(N^2) \).

To address the limitation of existing quantum self-attention models that neglect the intrinsic phase information in quantum states, Chen et al. \cite{chen2025quantum-c} proposed the Quantum Complex-Valued Self-Attention Model (QCSAM). The core insight of this method is to explicitly leverage the complex-valued similarity between quantum states to capture both amplitude and phase relationships, for which they introduced a Complex Linear Combination of Unitaries (CLCUs) framework that natively supports complex coefficients for weight aggregation. This model begins by encoding input data into corresponding quantum states, denoted as $|Q\rangle$, $|K\rangle$, and $|V\rangle$, via a PQC-based Quantum Feature Mapping module. The core of the model, the Quantum Complex Similarity Module, computes complex-valued attention weights by using an improved Hadamard test to compute the inner product between $|Q\rangle$ and $|K\rangle$ states, yielding a similarity weight that includes both real and complex parts. These complex weights and the quantum value states $|V\rangle$ are then processed by the novel Complex Linear Combination of Units (CLCUs) framework, which aggregates the value states to produce an output. Outputs from multiple patches are subsequently integrated using trainable CLCUs, and the resulting state is further refined by a Quantum Feedforward Network (QFFN) to enhance global context. Finally, the resulting quantum state is measured to complete the classification task.

QCSAM \cite{chen2025quantum-c} demonstrates superior performance compared to existing quantum self-attention models, achieving 100\% accuracy on MNIST and 99.2\% on Fashion-MNIST with just 4 qubits. It also shows strong scalability, with performance improving as qubit count increases for more complex multi-class tasks, and multi-head attention consistently outperforming single-head. Crucially, the novel use of complex-valued attention weights, derived from an improved Hadamard test, is key to its success, highlighting the importance of leveraging quantum phase information for enhanced self-attention mechanisms.

Shi et al. \cite{shi2023natural} proposed a quite straightforward quantum self-attention mechanism. They encode and organize the quantum circuit at the unit level of \( q_i \), \( k_j \), and \( v_j \), rather than structuring it at the level of \( Q \), \( K \), and \( V \) as done in other methods. Instead of explicitly computing attention scores, they apply PQC to each \( q_i \), \( k_j \), and \( v_j \) register to directly obtain the weighted value. The effectiveness of the model was validated on the small-scale datasets MC and RP. However, this method requires all \( q_i \), \( k_j \), and \( v_j \) pairs to be input at once, meaning that each token needs to be encoded for multiple times, resulting in a very high demand for qubit resources.

Overall, quantum pairwise self-attention methods implement token-wise pairwise similarity computation through quantum circuits, demonstrating the deep integration of quantum algorithms with Transformer models while also showcasing the diversity of quantum similarity interpretations. Among them, inner-product similarity methods tend to generate classical attention matrices (which, through conditional measurement control, can also technically associate attention coefficients with \( V \) in quantum states, such as in QKSAN \cite{zhao2024QKSAN}). The circuit execution complexity for these methods is \( O(N^2) \). In contrast, generalized similarity methods emphasize the uninterrupted execution on quantum computers. They adopt novel similarity metrics such as QLS or strongly entangled circuits, breaking free from the constraints of inner-product similarity in an attempt to obtain similarity information that is classically hard to simulate. However, these methods may still require \( O(N^2) \) complexity in terms of qubit count or circuit depth.

\begin{table*}[htbp]
\centering
\renewcommand{\arraystretch}{1.2} % Reduces vertical spacing (adjust value as needed)
\caption{Summary for the key features of the Quantum Holistic Attention method} \label{Summary for Quantum Holistic Attention}
\begin{tabular}{|m{1cm}|m{6cm}|m{5cm}|m{4cm}|}
\hline
\multicolumn{1}{|c|}{\textbf{Ref.}} & \multicolumn{1}{c|}{\textbf{Insight}} & \multicolumn{1}{c|}{\textbf{Contributions}} & \multicolumn{1}{c|}{\textbf{Limits}} \\ \hline
\cite{kerenidis2024quantum} & $\bullet$ Constructing quantum orthogonal circuits with RBS gates & $\bullet$ More stable training & $\bullet$ Complex classical pre-computation for RBS gate encoding \\ 
& $\bullet$ Directly apply quantum orthogonal layer to all tokens, generating global attention & $\bullet$ Sub-quadratic complexity & $\bullet$ Data loader requires deep circuits \\
      & $\bullet$ Develop data loader to encode entire matrix data & $\bullet$ Save qubit resources without adding to the measurement overhead &  \\ \hline
\cite{evans2024learning} & $\bullet$ Introduce QFT to transform token features to Fourier space for similarity calculation & $\bullet$ Sub-quadratic complexity & $\bullet$ Complex QFT circuits, difficult for NISQ devices \\ 
      &  &  & $\bullet$ Experimental validation is absent \\ \hline
\end{tabular}
\end{table*}

\subsubsection{Quantum Holistic Attention}\label{sec3.1.3}

Quantum Holistic Attention methods leverage quantum circuits to perform global mixing across all tokens simultaneously, eschewing the pairwise similarity computations of standard self-attention. Typically, they lack an explicit $QKV$ mapping process; instead, they integrate feature mapping and global token mixing within a single parameterized transformation, directly yielding the attention layer's output. Such approaches often involve specially designed global encoding strategies, enabling efficient information mixing with reduced computational complexity compared to pairwise attention methods. A summary table (Table \ref{Summary for Quantum Holistic Attention}) is provided to highlight the key features of all works in this category.

Kerenidis et al. \cite{kerenidis2024quantum} propose an innovative Quantum Vision Transformer, termed Quantum Compound Transformer (QCT), which is built upon two core components: Data Loader and Quantum Orthogonal Layer. The former employs unary amplitude encoding to efficiently transform a classical matrix \(\mathbf{X} \in \mathbb{R}^{N \times d}\) (e.g., an image divided into \( N \) patches, each with dimension \( d \)) into a quantum superposition state \( |\mathbf{X}\rangle = \frac{1}{\|\mathbf{X}\|} \sum_{i=1}^N \sum_{j=1}^d X_{ij} |\mathbf{e}_j\rangle |\mathbf{e}_i\rangle \). This process is realized by using two registers: an upper register with \( n \) qubits representing patch indices and a lower register with \( d \) qubits encoding the information of each patch. The Quantum Orthogonal Layer \cite{Kerenidis2021Classical}, implemented via parameterized Rotated Beam Splitter (RBS) gates, performs orthogonal matrix transformations, reducing the depth complexity of the parameterized circuit to \(\mathcal{O}(\log N)\), thereby balancing expressivity and hardware compatibility while mitigating the gradient vanishing issues common in variational circuits \cite{landman2022quantum}. 

Leveraging these tools, this work introduces a "compound" paradigm, where a second-order compound matrix \(\mathcal{V}_c^{(2)}\) (with dimension \(\binom{N+d}{2} \times \binom{N+d}{2}\)) integrates feature mapping and global weighting into a single high-order transformation to achieve global information mixing. First, Data Loader encodes the entire image into a quantum superposition state. Then, a single quantum orthogonal layer \(\mathbf{V}_c\) is applied across both registers, generating the output state \( |\mathbf{Y}\rangle = |\mathcal{V}_c^{(2)} X\rangle \), accomplishing global feature transformation and weighting in one step. The overall width and depth of the model are primarily determined by the Data Loader circuit, which are \( (N+d) \) and \(\mathcal{O}(\log N + 2N \log d)\), respectively. QCT \cite{kerenidis2024quantum} needs less quantum resource consumption than quantum pairwise attention methods. It harnesses quantum superposition and orthogonality to efficiently explore a larger Hilbert space while preserving gradient sharing among patches—akin to the global context of classical transformers but realized through quantum algorithms. Finally, by measuring the output state, the classical output patches \( (\mathbf{y_1}, \dots, \mathbf{y_N}) \in \mathbb{R}^{N \times d} \) are obtained. This method was validated on the MedMNIST dataset and achieved results outperforming classical benchmarks.

It is worth noting that, the spirit of QCT \cite{kerenidis2024quantum} is actually closer to MLP-Mixer \cite{tolstikhin2021mlp} than self-attention. MLP-Mixer employs alternating token-mixing MLP and channel-mixing MLP operations to directly perform global feature mixing across all patches. Similarly, the operation of the Quantum Compound Transformer follows this paradigm, replacing the classical MLP layers with quantum parameterized orthogonal layers. Its global transformation implicitly integrates the mixing of tokens and channels in a high-order transformation framework.

Evans et al. \cite{evans2024learning} propose a quantum self-attention variant based on Fourier transform and a kernel method. This work leverages the observation from FNet \cite{lee2021fnet} that unparameterized Fourier transforms can replace self-attention, maintaining high accuracy with significant training speedups, and extends this via the kernel convolution perspective \cite{guibas2021adaptive, pathak2022fourcastnet}, where self-attention can be represented by a convolution against a stationary kernel and can be simply computed in the Fourier domain. It implements this quantumly by globally encoding sequences into a tensor product quantum state \(|\psi\rangle = |\psi_1\rangle \otimes \cdots \otimes |\psi_N\rangle\), applying a Quantum Fourier Transform (QFT) to each token's state to shift into the frequency domain, followed by a variational kernel, \(U_{\text{kernel}}(\theta)\), for channel mixing and an inverse QFT to return to the computational basis, all within a single quantum circuit. A final variational unitary \(U_p(\theta)\) transfers information to a readout qubit for measurement. This reduces per-layer self-attention complexity from classical \(\mathcal{O}(N^2 d)\) to \(\mathcal{O}(N \log^2 d)\). Yet its limitations include the stationary kernel assumption, which underpins the convolution reformulation but is not inherently valid for self-attention due to its dynamic, context-dependent weights rather than sole reliance on relative positions, risking reduced expressive power for complex, non-stationary tasks. Additionally, it fails to give publicly available experimental data to validate its feasibility.

\subsubsection{Quantum-Assisted Optimization}\label{sec3.1.4}

Instead of replacing classical Transformer components with quantum circuits, this category focuses on structurally optimizing attention matrices—through sparsification for efficiency or doubly stochastic normalization for improved regularization—by leveraging quantum algorithms. A summary table (Table \ref{Summary for Quantum-Assisted Optimization}) is provided to highlight the key features of all works in this category.

\begin{table*}[htbp]
\centering
\renewcommand{\arraystretch}{1.2} % Reduces vertical spacing (adjust value as needed)
\caption{Summary for the key features of the Quantum-Assisted Optimization method} \label{Summary for Quantum-Assisted Optimization}
\begin{tabular}{|m{1cm}|m{6cm}|m{5cm}|m{4cm}|}
\hline
\multicolumn{1}{|c|}{\textbf{Ref.}} & \multicolumn{1}{c|}{\textbf{Insight}} & \multicolumn{1}{c|}{\textbf{Contributions}} & \multicolumn{1}{c|}{\textbf{Limits}} \\ \hline
\cite{gao2023fast} & $\bullet$ Employing Grover's algorithm for attention matrix sparsification & $\bullet$ Accelerating sparsification from $O(N)$ to $O(\sqrt{N})$ & $\bullet$ Oracle is needed, posing high difficulty for NISQ devices \\ 
&  &  & The sparsity assumption may not always hold \\ \hline
\cite{born2025quantum} & $\bullet$ Applying specialized quantum circuits for attention matrix double stochasticity & $\bullet$ Yielding quantum inductive bias without classical analogue & $\bullet$ No reduction in classical computational load or complexity \\ 
      &  & $\bullet$ Superior performance to classical double stochasticization &  \\ \hline
\end{tabular}
\end{table*}

\textbf{A. Sparsification of the attention matrix}

Gao et al. \cite{gao2023fast} propose a model optimization strategy for attention computation in large language models, inspired by the observation that attention matrices are often sparse \cite{child2019generating, zhang2023h2o}. This approach leverages Grover's Search, a quantum algorithm that efficiently finds \(k\) target elements in an unstructured \(N\)-element set in \(\tilde{O}(\sqrt{Nk})\) time by exploiting superposition and interference, offering a quadratic speedup over classical \(\mathcal{O}(Nk)\) search. Motivated by this, the method uses Grover's Search to accelerate identification of sparse, significant entries in attention matrices, classically encodes \(Q\) and \(K\) matrices, locates \(k\) entries per row exceeding a threshold \(\tau\) in \(\tilde{O}(\sqrt{Nk} d)\) time, and constructs a sparse matrix \(B\) with a rank-1 component through classical operations, reducing inference complexity from \(\mathcal{O}(N^2 d)\) to \(\tilde{O}(N^{1.5} k^{0.5} d + Nkd)\). By limiting quantum usage to search acceleration, it achieves polynomial speedup with error bounds of \(O(\eta^2)\) when sparsity (\(k \ll N\)) holds. However, the effectiveness of this method relies on the \((\tau, k)\)-good sparsity assumption and efficient oracle access. It lacks experimental validation to confirm its performance in practical applications.

\textbf{B. Doubly stochastic normalization of the attention matrix}

The softmax-based attention mechanism in classical Transformers normalizes the attention matrix such that each row sums to 1, forming a right stochastic matrix. While effective in many settings, this normalization often leads to overly concentrated attention, entropy collapse, and unstable training dynamics, particularly in small-sample learning or multimodal tasks. To address these issues, prior work such as Sinkformer \cite{sander2022sinkformers} introduced the Sinkhorn algorithm, which iteratively normalizes the attention matrix into a Doubly Stochastic Matrix (DSM)—one where both rows and columns sum to 1—resulting in a more “democratic” attention distribution and better information coverage. However, Sinkhorn’s algorithm is iterative, approximative, non-parametric and thus inflexible w.r.t. the obtained doubly stochastic matrix (DSM).

To overcome this limitation, Born et al. \cite{born2025quantum} propose a hybrid classical-quantum doubly stochastic Transformer (QDSFormer) that replaces the softmax in the self-attention layer with a PQC. This method is built upon a theoretical foundation: DSMs can be obtained with a PQC, yielding a novel quantum inductive bias for DSMs with no known classical analogue \cite{mariella2024quantum}. Specifically, the QDSFormer produces a unitary matrix $\mathbf{U}(p; \theta)$, from which a non-negative matrix is constructed through the Hadamard (i.e., element-wise) product with its complex conjugate $\bar{\mathbf{U}}$:
$$
\mathbf{A} = \mathbf{U}(p; \theta) \circ \bar{\mathbf{U}}(p; \theta).
$$
This matrix is subsequently normalized to approximate a DSM and used to replace the classical softmax attention matrix in the Transformer. Unlike classical methods such as Sinkhorn, The doubly stochasticization process of the attention matrix in QDSFormer is parametric, and its parameters can be jointly optimized during training.

The authors conduct simulation-based evaluations using Qiskit \cite{Qiskit}, and benchmark QDSFormer on MNIST, FashionMNIST, MedMNIST and Eureka datasets. Experimental results demonstrate that QDSFormer \cite{mariella2024quantum} consistently outperforms ViT \cite{vaswani2017attention} and Sinkformer \cite{sander2022sinkformers} in terms of accuracy, information retention, entropy, training stability, and generalization. This work presents the first demonstration of quantum circuits as a means to construct learnable attention distributions, introducing a novel quantum inductive bias into Transformer architectures.

This work represents the first demonstration of quantum circuits being used to construct learnable attention distributions, introducing a novel quantum inductive bias into Transformer-based architectures and opening new avenues for trainable quantum attention mechanisms.

%%%%%%%%%%%%%%%%%%%%%%%%%%%%%%%%%%%%%%%%%%%%%%%%%%%%%%%%%%%%%%%%%%%%%%%%%%%%%%%%%%%%%%%%%%%%%%%%%%%%%

\subsection{QLA-based Quantum Transformers}\label{sec3.2}

Transformer models based on QLA offers a promising direction, although research in this area remains largely theoretical. Their related approaches are mainly designed for the future era of fault-tolerant quantum computing, where the powerful tools of quantum linear algebra may enable the exponential acceleration of quantum Transformers. We provide a brief introduction to existing QLA-based Transformer models.

Quantum linear algebra methods leverage the unique properties of quantum computing to efficiently solve fundamental problems in classical linear algebra, enabling exponential speedup for certain tasks. Since the core computations of the Transformer model rely heavily on matrix operations, quantum Transformers based on quantum linear algebra have emerged as a novel approach for quantumizing Transformer models. The core idea of QLA-based quantum transformer models is to leverage block encoding for matrix operations while utilizing QSVT for implementation of non-linear transformation (e.g., softmax or GELU functions). Meanwhile, other techniques such as Linear Combination of Unitaries (LCU) \cite{childs2012hamiltonian} and amplitude transformation are incorporated to implement functions like residual connections and layer normalization.

Guo et al. \cite{guo2024quantum} present a pioneering QLA-based quantum Transformer architecture. For the given pre-trained parameters, this work utilities quantum means to implement all the steps in a transformer block for the inference stage. It also analyzes the complexity of each quantum subroutine and the overall complexity of a transformer block.

First, for the given pre-trained weights, matrix \( QK^T \) is directly obtained, and then is encoded in Quantum circuit by block encoding. Then, QSVT is applied to perform an element-wise matrix function and then implement the softmax function by polynomial approximation. Immediately following this, the multiplication of the attention matrix and $V$ can be conveniently implemented within the block-encoding framework. Besides, residual connections are realized by linear combination of block encodings, while layer normalization employs amplitude transformations to standardize vectors, both integrated seamlessly with quantum states’ inherent normalization. Regarding FFN, the block-encoding technique is again employed to encode the parameter matrices of the linear layers, while QSVT is applied to implement the GELU nonlinear activation function.

This method achieves a complexity of $\mathcal{O}(d n^2 \alpha^2 \log^2\left( \frac{1}{\epsilon} \right))$ for a single-layer output state preparation (where \( \tilde{d} \) is the embedding dimension, \( n = \log N \) with \( N \) as sequence length, \( \alpha \) as normalization factors, and \( \epsilon \) as error), offering potential exponential speedups over classical $\mathcal{O}(N^2 d + N d^2)$ complexity. 

However, it is noteworthy that the parameters of this model are pre-trained (may have been trained by a classical model, but the paper does not explicitly state it), meaning that the input data for block encoding \(QK^T\) is fixed. This highlights a limitation of the QLA-based approaches: since block encoding requires a specific unitary (denotes as $U_A$) for a given matrix $A$ (see Eq. \ref{block-encoding}), updating $A$ necessitates recalculating the $U_A$ matrix each time, which is computationally intensive, potentially offsetting the inherent quantum acceleration advantages provided by QLA techniques.

Another study \cite{liao2024gpt} adopts a hybrid approach combining quantum linear algebra with variational quantum algorithms and provides a specific structure for the quantum circuit structure. This approach uses block encoding for the attention matrix \( QK^T \), variational quantum circuits for \( W_V \), and matrix vectorization for their multiplication, yielding quantum self-attention results while omitting softmax to reduce the complexity. The residual connection employs a Hadamard gate on an ancillary qubit for superposition, controlled operations to link self-attention output and input \( X \), followed by another Hadamard gate and post-selection measurement. FFN is implemented in two steps: parallel swap tests compute inner products, amplitude estimation stores results, phase estimation generates binary representations, and an arithmetic circuit computes ReLU. Compared to \cite{guo2024quantum}'s theoretical framework, this approach is concrete, integrating linear algebra and variational algorithms, with \( QK^T \) given directly, while \( W_V \) and FFN weights as variational parameters. One limitation of this approach \cite{liao2024gpt} is that it requires intermediate quantum measurements between layers—each block’s output must be measured and converted to classical data before serving as input to the next block.

Khatri et al. \cite{khatri2024quixer} propose a quantum self-attention variant based on Linear Combination of Unitaries (LCU) \cite{childs2012hamiltonian} and Quantum Singular Value Transformation (QSVT). This work discards the traditional QKV structure by directly encoding input embeddings via LCU into unitary matrices, forming a superposition of all tokens. QSVT extracts polynomials (typically second-order) capturing all token-pair interactions, with measurement results serving as quantum self-attention outputs, simplifying the quantum linear algebra-based model for basic numerical simulations. Another study \cite{xue2024end} focuses on implementing quantum residual connections by using Quantum Random Access Memory (qRAM) \cite{giovannetti2008qRAM}, thus not requiring uninterrupted quantum evolution, while allowing quantum-classical data conversion interfaces. However, the physical realization of qRAM may be more challenging than fault-tolerant quantum computers, potentially requiring decades of hardware advancements.

In the NISQ era, certain steps in PQC-based models—such as softmax, residual connections, and layer normalization—are typically implemented classically or omitted entirely as shown in Fig. \ref{roadmap}. By contrast, in the fault-tolerant quantum computing era, researchers can leverage a richer set of quantum linear algebra tools—thanks to significantly increased qubit counts and improved gate fidelities—to implement a complete quantum transformer structure, theoretically offering stronger potential for quantum acceleration and enhancement.

%%%%%%%%%%%%%%%%%%%%%%%%%%%%%%%%%%%%%%%%%%%%%%%%%%%%%%%%%%%%%%%%%%%%%%%%%%%%%%%%%%%%%%%%%%%%%%%%%%%%%%%%%%%%%%%%%%%%%%%%%

\section{Preliminary Validation and Applications of Quantum Advantage}\label{sec4}

\subsection{Better Performance on Small-Scale Datasets}

\begin{table*}[h]
\centering
\begin{threeparttable}
\caption{Binary classification accuracy of some models on NLP datasets. The numerical precision of the results is retained as in the original references. The best result in each column is highlighted in bold.}
\label{NLP results}
\begin{tabular}{C{3cm}C{2cm}C{2cm}C{2cm}C{2cm}C{2cm}}
\toprule%
\textbf{Model} & \textbf{MC} & \textbf{RP} & \textbf{Yelp} & \textbf{IMDb} & \textbf{Amazon} \\ \midrule
CSANN & —— & —— & 83.11\% & 79.67\% & 83.22\% \\
DisCoCat & 79.8\% & 72.3\% & —— & —— & —— \\ \midrule
\cite{li2024quantum} & \textbf{100\%} & 67.74\% & 84.79\% & 80.28\% & 84.25\% \\
\cite{chen2025quantum} & \textbf{100\%} & 75.63\% & \textbf{84.96\%} & \textbf{84.82\%} & \textbf{87.48\%} \\
\cite{zheng2023Design} & \textbf{100\%} & \textbf{87.10\%} & —— & —— & ——  \\
\cite{shi2023natural} & \textbf{100\%} & 74.19\% & —— & —— & ——  \\
\cite{wei2023povm} & \textbf{100\%} & 77.42\% & —— & —— & ——  \\ \bottomrule
\end{tabular}
\begin{tablenotes}
    \footnotesize
    \item '——' indicates that there is no experiment conducted.
\end{tablenotes}
\end{threeparttable}
\end{table*}

\begin{table*}[h]
\centering
\begin{threeparttable}
\caption{Binary classification accuracy of some models on NLP datasets. The numerical precision of the results is retained as in the original references. The best result in each column is highlighted in bold.}
\label{CV results}
\begin{tabular}{C{1.6cm}C{1.8cm}C{1.8cm}C{1.8cm}C{1.8cm}C{1.8cm}C{1.8cm}C{1.8cm}}
\toprule%
\textbf{Model} & \textbf{MNIST(0/1)} & \textbf{Fashion-MNIST(0/1)} & \textbf{CIFAR-10(0/1)} & \textbf{Mini-Imagenet(0/1)} & \textbf{DermaMNIST*} & \textbf{BreastMNIST*} & \textbf{RetinaMNIST*} \\ \midrule
CSANN \cite{zhang2024light-weight} & 99.69\% & 99.13\% & 82.50\% & 86.50\% & —— & —— & —— \\
ViT \cite{dosovitskiy2020vit} & —— & —— & —— & —— & 72.7\% & 83.3\% & 54.8\% \\ \midrule
\cite{zhao2024QKSAN} & 99\% & 98.05\% & —— & —— & —— & —— & —— \\
\cite{shi2024QSAN} & \textbf{100\%} & —— & 86.67\% & —— & —— & —— & —— \\
\cite{zhang2024light-weight} & 99.91\% & 99.06\% & 87.36\% & 86.75\% & —— & —— & —— \\
\cite{zhang2025hqvit} & \textbf{100\%} & —— & \textbf{88.5\%} & \textbf{88.5\%} & —— & —— & —— \\ 
\cite{chen2025quantum-c} & \textbf{100\%} & \textbf{99.2\%} & —— & —— & —— & —— & —— \\
\cite{kerenidis2024quantum} & —— & —— & —— & —— & \textbf{73.4\%} & \textbf{84.6\%} & \textbf{56.5\%} \\ \bottomrule
\end{tabular}
\begin{tablenotes}
    \footnotesize
    \item '——' indicates that there is no experiment conducted.
    \item The data in the table retains the original precision as presented in the source.
    \item Datasets marked with an '*' are subsets from the MedMNIST dataset collection. The original study conducted experiments on 12 subsets in total, and we present the results for three of them. It is important to note that quantum models do not outperform classical baselines on all of these subsets. 
\end{tablenotes}
\end{threeparttable}
\end{table*}

Current PQC-based quantum Transformers have shown promising potential on small-scale problems, largely driven by strategic hybrid quantum-classical architectures. These designs integrate quantum modules with essential classical components to leverage the strengths of both paradigms.

As shown in Table~\ref{NLP results}, quantum-enhanced models achieve competitive performance on small NLP datasets. For example, \cite{chen2025quantum} report 84.96\% and 87.48\% accuracy on the Yelp and Amazon sentiment datasets, respectively; \cite{zheng2023Design} achieve 87.10\% on RP; and all models reach 100\% on MC. While full classical comparisons are not always available, these results highlight the viability of quantum methods for specific NLP tasks.

In CV tasks (Table~\ref{CV results}), similar trends are observed. \cite{zhang2025hqvit} report 88.5\% on CIFAR-10(0/1) and Mini-Imagenet(0/1); \cite{chen2025quantum-c} reach 99.06\% on Fashion-MNIST(0/1); and \cite{kerenidis2024quantum} achieve 73.4\% and 84.6\% on DermaMNIST and BreastMNIST, respectively. These results underscore the capacity of quantum-enhanced approaches to contribute to visual pattern recognition.

\begin{figure}[h]
    \centering
    \includegraphics[width=0.49\textwidth]{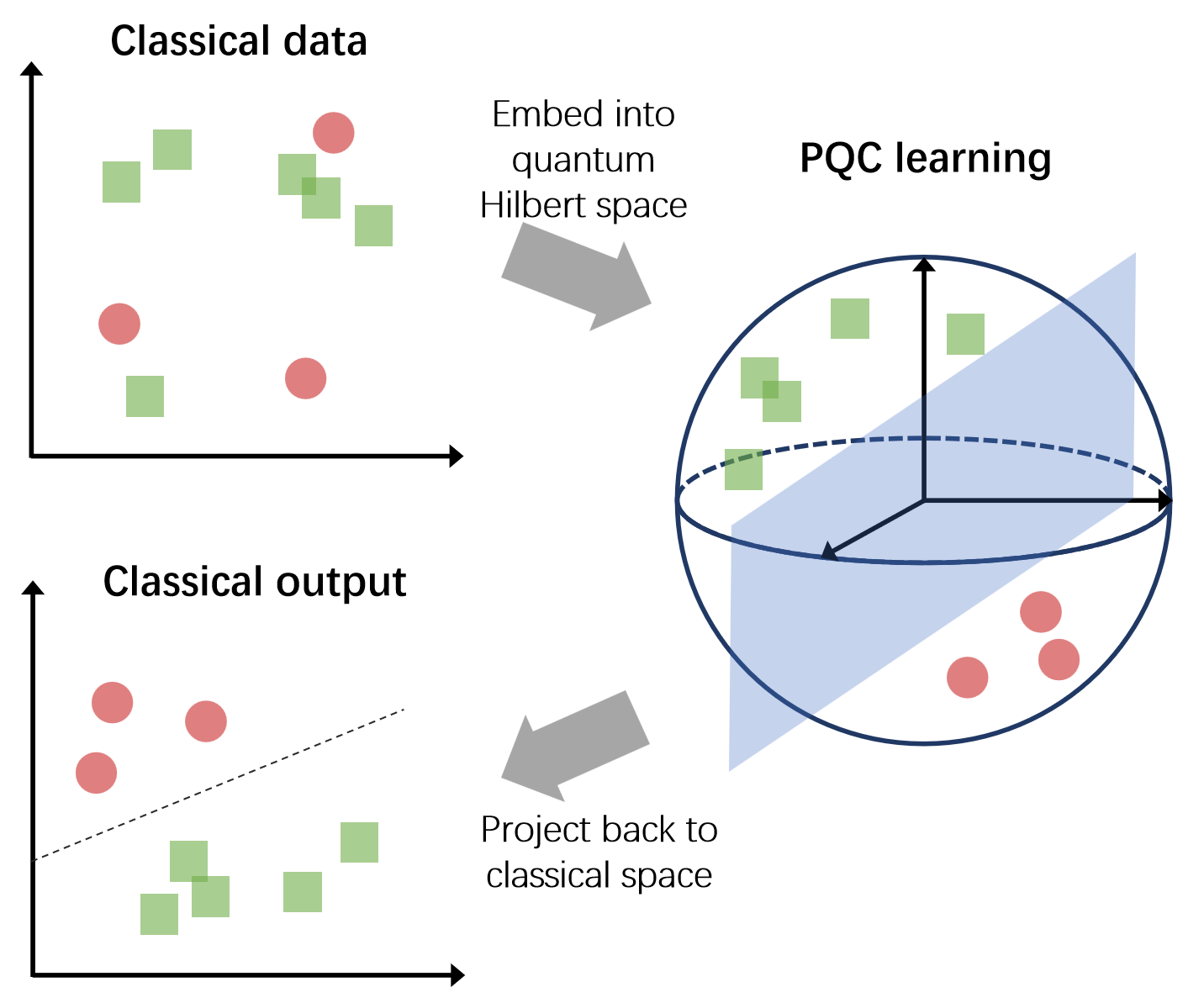}
    \caption{Schematic Representation of the Underlying Principle of PQC Learning. The quantum Hilbert space, inherently a high-dimensional space, is simplified here by using a Bloch sphere. This figure depicts the process of embedding classical data (represented by green squares and red circles) into the quantum Hilbert space, followed by projection back to classical space for output, highlighting the transformation enabled by PQC learning \cite{houssein2022machine}.} \label{PQC learning}
\end{figure}

These performance gains are primarily attributed to the enhanced representation capabilities offered by the high-dimensional quantum space. Quantum circuits can map low-dimensional classical data into an exponentially larger Hilbert space, as shown in Fig. \ref{PQC learning}. Within this vast space, PQC can perform complex, non-linear unitary transformations, which inherently provide a richer and more intricate feature representation than classical linear or shallow non-linear operations. This may allow quantum models to capture subtle correlations and patterns within the data that might be challenging for classical models of comparable size or complexity.

\subsection{Potential in Real-World Applications}

\begin{table*}[h]
\centering
\begin{threeparttable}
\caption{Performance on Real-World Application. In the 'Category' column, the Roman numeral ‘\textbf{I}’ indicates \textit{QKV-only quantum mapping} models, and ‘\textbf{II}’ indicates \textit{Quantum pairwise attention} models. In the 'Result' column, 'C' denotes the performance of classical benchmarks, while 'Q' represents the results of the quantum Transformer model. Bold text indicates better performance. It is important to note that most works conduct experiments on multiple datasets and across various metrics; here, we have only selected the most representative results, with more necessary details explained in the footnotes.}
\label{RW results}
\begin{tabular}{C{0.8cm}C{0.8cm}C{3cm}C{4.8cm}C{3cm}C{2cm}}
\toprule%
\textbf{Model} & \textbf{Category} & \textbf{Application scenario} & \textbf{Specific task} & \textbf{Evaluation metrics} & \textbf{Result(C/Q)} \\ \midrule
\cite{kamata2025molecular} & \textbf{II} & Chemistry & Estimation of the ground-state energies of molecules & Average ground-state energy estimation error ($H_2$) & 9.3e-3/\textbf{8.9e-3} \\ \midrule
\cite{smaldone2025hybrid} & \textbf{II} & Chemistry & Molecule configurations generation & Accuracy of tokens prediction & 0.616/\textbf{0.620} \\ \midrule
\cite{unlu2024hybrid} & \textbf{I} & High-Energy Physics & Distinguishing photons and electrons through images & Accuracy of binary classification & \textbf{0.718}/$\textbf{0.718}^1$ \\ \midrule
\cite{cara2024quantum} & \textbf{I} & High-Energy Physics & Distinguishing quark jets and gluon jets through images & Accuracy of binary classification & \textbf{0.798}/0.780 \\ \midrule
\cite{tesi2024quantum} & \textbf{II} & High-energy physics & Distinguishing quark jets and gluon jets through images & Accuracy of binary classification & \textbf{0.679}/0.676 \\ \midrule
\cite{dutta2024aq} & \textbf{I} & Fluid Dynamics & Predicting the numerical solution of the Navier-Stokes equations & Test loss & 0.0631/\textbf{0.0593} \\ \midrule
\cite{dutta2024qadqn} & \textbf{I} & Finance & Optimizing trading strategies on historical S \& P 500 \cite{spglobal2024} index data & Return(\%) & $74.42^2$/\textbf{78.91} \\ \midrule
\cite{dutta2024qadqn} & \textbf{I} & Time Series Forecasting & Electricity Load Forecasting$^3$ & Mean Absolute Error & $0.413^4$/\textbf{0.407} \\
\bottomrule
\end{tabular}
\begin{tablenotes}
    \footnotesize
    \item 1: This work experiments with several different quantum Transformer model variants i.e., Column Max, and we select the one with the highest accuracy for this metric.
    \item 2: The classic baseline in this work is not an equivalent counterpart to the quantum transformer; rather, it uses Buy \& Hold \cite{bodie2018investments}, a non-intelligent strategy that serves as the most fundamental benchmark in the financial domain.
    \item 3: This work conducted experiments on two datasets for this task, and we selected ETTh2.
    \item 4: This work compares against multiple classical benchmark, and we selected iTransformer, which performed best among all classical benchmark for this metric.
\end{tablenotes}
\end{threeparttable}
\end{table*}

In the aforementioned studies on computer vision and natural language processing, quantum Transformer models have primarily been evaluated on small-scale public datasets to demonstrate methodological feasibility and performance. In contrast, other applications—such as in chemistry, high-energy physics, and finance—aim to employ quantum Transformers for solving real-world, domain-specific problems. We refer to these cases collectively as real-world applications, as shown in Table \ref{RW results}. 

These applications leverage hybrid quantum models to address challenges in domains like chemistry, high-energy physics, fluid dynamics, finance and time series forecasting. In chemistry, \cite{smaldone2025hybrid} and \cite{kamata2025molecular} investigate for problems such as the estimation of ground-state energies of molecules and molecule configuration generation. These works reported benefits including lower estimation error with stronger generalization and improved accuracy and validity in generation tasks. In high-energy physics, \cite{cara2024quantum}, \cite{unlu2024hybrid}, and \cite{tesi2024quantum} were applied to tasks involving distinguishing particles and jets through images. Although the performance of quantum Transformer models is currently slightly below that of classical ViT models, the results are already quite comparable. these models \cite{cara2024quantum, unlu2024hybrid, tesi2024quantum} still have considerable room for improvement in terms of architecture design and the selection of PQCs, suggesting that holding strong potential for further performance gains. In applications such as fluid dynamics \cite{dutta2024aq}, finance \cite{dutta2024qadqn}, and time-series forecasting \cite{chakraborty2025integrating}, quantum Transformers are applied to regression tasks, where they have demonstrated superior performance over their classical counterparts in terms of prediction error.

Furthermore, as shown in Table \ref{RW results}, all real-world quantum Transformer applications are carried out by models that belong to either the `QKV-only quantum mapping` or the `Quantum pairwise attention` categories. All of these models adopt a hybrid quantum–classical architecture. In particular, those `QKV-only quantum mapping` models incorporate the largest proportion of classical components among all categories, which may explain their prevalence in real-world tasks. The hybrid model approach is crucial here, as it allows the quantum module to handle the “quantum-friendly” core computations, while classical components manage data input/output and the overall algorithmic flow, making these applications feasible on current hardware. This suggests that, in the NISQ era, models with a higher classical-to-quantum ratio remain the most practical and promising direction for deploying quantum Transformers in real-world tasks.

%%%%%%%%%%%%%%%%%%%%%%%%%%%%%%%%%%%%%%%%%%%%%%%%%%%%%%%%%%%%%%%%%%%%%%%%%%%%%%%%%%%%%%%%%%%%%%%%%%%%%%%%%%%%%%%%%%%%%%%%%%%%

\section{Challenges and Potential Solutions}\label{sec5}

While the preceding sections have highlighted promising preliminary validations of quantum advantage in Transformer models, it is equally crucial to acknowledge the significant challenges that impede their broader development and real-world deployment. These challenges are multifaceted, stemming from fundamental limitations of current quantum hardware (NISQ devices), the inherent properties of variational quantum algorithms, and the immaturity of quantum machine learning theory. Effectively addressing these hurdles is paramount for unlocking the full potential of quantum Transformers and moving beyond their current constrained applications.

\subsection{Challenges in PQC-based Quantum Transformers}\label{sec4.1}

\subsubsection{Complexity Trade-offs} \label{sec4.1.1}

\textbf{A. Current challenges}

\begin{table*}[h]
\centering
\begin{threeparttable}
\caption{Evaluation of the quantum resource requirements, where $N$ denotes the sequence length, and $d$ represents the feature dimension of the token. Only key method-category papers are listed in this table. Works on \textit{Method Improvements} or \textit{pplication-Oriented} are omitted, as they do not affect the quantum resource complexity of the original models in big-$\mathcal{O}$ terms.}
\label{Quantum resource}
\begin{tabular}{C{1.4cm}C{0.7cm}C{1.8cm}C{2cm}C{1.6cm}C{1.3cm}C{1.5cm}C{2.2cm}C{1.5cm}}
\toprule%
\textbf{Category} & \textbf{Model} & \textbf{\#Qubits} & \textbf{Circuit depth} & \textbf{\#Measurement per time} & \textbf{Circuit execution times} & \textbf{\#Total measurement} & \textbf{Outputs of quantum circuits} & \textbf{Remaining classical complexity} \\ \midrule
QKV-only Quantum mapping & \cite{li2024quantum} & $O(d)$ & $O(\mathrm{poly}(d))$ & $O(d)$ & $O(N)$ & $O(Nd)$ & $\{\textbf{q}_i\}_{i=1}^N$, $\{\textbf{k}_i\}_{i=1}^N$, $\{\textbf{v}_i\}_{i=1}^N$ & $O(N^2d)$ \\ \midrule
\multirow{5}{*}{\parbox{1.5cm}{\centering Quantum\\pairwise\\attention}} & \cite{chen2025quantum} & $O(d)$ & $O(\mathrm{poly}(d))$ & $O(d)$ & $O(N^2)$ & $O(N^2d)$ & $\{A_{ij}\}_{i,j=1}^N$, $\{\textbf{v}_i\}_{i=1}^N$ & $O(N^2d)$ \\ \cmidrule{2-9}
& \cite{zhang2025hqvit} & $O(\log(Nd))$ & $O(\mathrm{poly}(\log d))^*$ & $O(\log d)$ & $O(N^2)$ & $O(N^2 \log d)^{**}$ & $\{A_{ij}\}_{i,j=1}^N$, $\{\textbf{v}_i\}_{i=1}^N$ & $O(N^2d)$ \\ \cmidrule{2-9}
& \cite{smaldone2025hybrid} & $O(d)$ & $O(\mathrm{poly}(d))$ & $O(d)$ & $O(N^2)$ & $O(N^2d)$ & $\{A_{ij}\}_{i,j=1}^N$, $\{\textbf{v}_i\}_{i=1}^N$ & $O(N^2d)$ \\ \cmidrule{2-9}
& \cite{zhao2024QKSAN}$^1$ & $O(d)$ & $O(\mathrm{poly}(d))$ & $O(d)$ & $O(N^2)$ & $O(N^2d)^{**}$ & $\{\textbf{y}_i\}_{i=1}^N$ & $O(Nd)$ \\ \cmidrule{2-9}
& \cite{shi2024QSAN} & $O(N \log d + N^2)$ & $O(N^2 \log d)^*$ & $O(\log d)$ & $O(1)$ & $O(\log d)$ & {\textbf{y}} & $O(d)$ \\ \cmidrule{2-9}
& \cite{zheng2023Design} & $O(N \log d)$ & $O(N^2 \mathrm{poly}(\log d))^*$ & $O(1)$ & $O(1)$ & $O(1)$ & Probability of the predicted class & —— \\ \midrule
\multirow{2}{*}{\parbox{1.5cm}{\centering Quantum\\global\\attention}} & \cite{kerenidis2024quantum} & $O(N + d)$ & $O(N \log d)$ & $O(d)$ & $O(N)$ & $O(Nd)^{**}$ & $\{\textbf{y}_i\}_{i=1}^N$ & $O(Nd)$ \\ \cmidrule{2-9}
& \cite{evans2024learning} & $O(N \log d)$ & $O(N \log^2 d + \mathrm{poly}(N \log d))^*$ & $O(1)$ & $O(1)$ & $O(1)$ & Probability of the predicted class & —— \\ \bottomrule
\end{tabular}
\begin{tablenotes}
    \footnotesize
    \item 1: This method has been implemented by using both amplitude encoding and qubit encoding. Here we select the case of qubit encoding.
    \item '——' indicates that there is no classical computational overhead.
    \item '*' indicates that this method uses amplitude encoding, which may significantly increases the circuit depth.
    \item '**' indicates that the measurement process of this method includes post-selection measurements, which will significantly increases the number of sampling required.
\end{tablenotes}
\end{threeparttable}
\end{table*}

Reducing the $O(N^2 d)$ complexity of the attention mechanism—particularly the cost associated with computing all pairwise token interactions—is a core motivation behind PQC-based quantum Transformer models. A typical approach is to offload parts of this computation, such as $QKV$ projections or similarity estimation, from classical processing to quantum subroutines. In some cases, methods like the Swap Test or Hadamard Test allow inner-product estimation in $O(1)$ quantum time per token pair, thus offering a potential local quantum speedup. 

However, achieving reductions in classical complexity does not equate lowering the overall computational cost. In practice, reducing classical computation almost always entails a corresponding increase in quantum resource consumption. Table \ref{Quantum resource} provides a comprehensive analysis of quantum resource requirements across a variety of PQC-based Transformer models. The selected metrics are explained as follows:

- \textit{Number of qubits (width).} This reflects the spatial complexity. Given the limited number of qubits in NISQ devices, all methods adopt the assumption of qubit reuse (i.e., serial execution) rather than parallel computation.

- \textit{Circuit depth}: This reflects the time complexity of quantum computation. Here, we consider the depth of PQC as the primary measure, typically expressed as \( O(\text{poly}(n)) \), where \( n \) is the number of qubits. Multi-qubit gates (e.g., CNOT) are treated as single-depth units without gate decomposition. Notably, the reported depth excludes additional overhead from data encoding—such as amplitude encoding—which is marked with an asterisk (*).

- \textit{Measurement Overhead.} This represents the overhead of quantum-classical interactions and is a key factor in the actual cost of running quantum algorithms. The measurement overhead per execution depends on the required feature dimensions (e.g., extracting a \( d \)-dimensional vector requires \( O(d) \) measurements). The total measurement overhead is the product of the measurement overhead per execution and the number of circuit executions. Notably, to obtain probabilistic results, multiple repeated samplings are required in actual experiments, but in this paper, we provide only qualitative annotations rather than quantitative analysis.

- \textit{Number of circuit executions.} Different methods invoke quantum circuits at varying frequencies, affecting overall computational cost. Under the qubit reuse assumption, the QKV-only method requires token-by-token mapping. Thus it needs to run the circuit $N$ times, whereas the inner product attention method often requires $N^2$ runs (because each time only one attention score is obtained).

- \textit{Remaining classical computational complexity.} It captures any residual classical cost, since most quantum Transformer models partially rely on classical computations (e.g., FFN or post-processing).

We can draw some remarkable insights from Table \ref{Quantum resource}:

1) \textit{Persistent Quadratic Complexity of QKV-Only Quantum mapping and Quantum Pairwise Attention Methods.} The QKV-only Quantum mapping and quantum pairwise attention methods largely fail to achieve complexity below the quadratic scaling with sequence length \( N \). For instance, the QKV-only Quantum mapping method, such as \cite{li2024quantum}, relies on classical algorithms for core self-attention computation, resulting in remaining classical computational overhead of \( O(N^2 d) \). Similarly, quantum pairwise attention methods \cite{chen2025quantum}, \cite{zhang2025hqvit}, \cite{smaldone2025hybrid}, \cite{zhao2024QKSAN} exhibit \( O(N^2 d) \) complexity in at least one dimension, requiring \( O(N^2) \) circuit runs and yielding total measurement counts of \( O(N^2 d) \) or \( O(N^2 \log d) \). Additionally, \cite{shi2024QSAN} and \cite{zheng2023Design} demonstrate \( O(N^2) \) complexity in terms of circuit width or depth, indicating no significant reduction from the classical \( O(N^2) \) bottleneck.

2) \textit{Sub-Quadr Complexity of Quantum Holistic Attention Methods with Limitations.} Quantum Holistic Attention methods appear to reduce overall complexity below \( O(N^2) \). For example, \cite{kerenidis2024quantum} achieves a width of \( O(N + d) \), a depth of \( O(N \log d) \), and total measurement counts of \( O(Nd) \), while \cite{evans2024learning} reports a width of \( O(N \log d) \), a depth of \( O(N \log^2 d + \text{poly}(N \log d)) \), and total measurement counts of \( O(1) \). However, it is critical to note that these methods deviate from the original Transformer based on inner-product similarity, aligning more closely with variants such as MLP-Mixer \cite{tolstikhin2021mlp} or Fourier Transformers \cite{lee2021fnet}, which inherently exhibit linear complexity with \( N \) in their classical forms. Thus, the complexity reduction may not reflect a genuine quantum advantage over the analogous classical baseline.

3) \textit{Complexity is Redistributed but Not Removed.} Fundamentally, no current PQC-based Transformer model fundamentally breaks the $O(N^2)$ barrier in attention computation. Some shift classical cost to the quantum computing domain, but often at the expense of measurement repetition, circuit depth, or ancilla qubit counts. This pattern reflects a fundamental constraint in PQC-based Transformers: complexity is redistributed rather than eliminated. To reduce classical complexity, one must accept increased quantum cost—often without a net reduction in total computational burden.

\textbf{B. Potential solutions}

Whether current VQA models can achieve a significant improvement in computational complexity for general machine learning tasks remains an open question. Therefore, whether PQC-based quantum Transformers have the potential to overcome the quadratic complexity barrier—or whether there exist inherent limitations—still requires further exploration.

Nevertheless, we can shift our attention to quantumizing existing classical Transformer variants that already exhibit sub-quadratic complexity. Preliminary investigations have been conducted in this direction using Quantum Holistic Attention mechanisms. Future research could focus on quantum adaptations of Transformer variants such as Performer \cite{choromanski2020rethinking} or Linformer \cite{wang2020linformer}, both of which inherently operate with sub-quadratic complexity. This approach could help reduce both classical and quantum resource requirements.

When these models are integrated with efficient encoding schemes and lightweight PQC modules, it may become possible to retain the expressive power of attention mechanisms while achieving greater scalability. Ensuring architectural compatibility with fundamental classical components such as FFN and residual blocks remains essential. At the same time, enhancing the theoretical interpretability of quantum attention mechanisms is also crucial.

\subsubsection{Scalability and Generalization} \label{sec4.1.2}

\textbf{A. Current challenges}

While quantum Transformers have shown promising early results, these findings should be interpreted with caution, as several existing limitations continue to pose challenges to further advancement.

1) The scope of currently demonstrated quantum advantage remains highly restricted. Nearly all existing experiments are confined to small-scale datasets with reduced input dimensionality, typically in binary or few-class classification settings. Consequently, the observed performance gains are predominantly localized, raising significant concerns about the extensibility of these models to larger, more complex real-world applications (e.g., full ImageNet, large-scale NLP tasks) that demand genuine scalability.

2) Methodological inconsistencies severely undermine the credibility of generalization claims. Common practices include the use of classical baselines that are structurally weakened to match the constraints of quantum model design, which can artificially inflate the perceived performance gains of the quantum model. Furthermore, diverse preprocessing strategies across different studies (e.g., PCA, classical projection layers, truncated embeddings) introduce significant variability, obscuring the true attribution of performance improvements and making fair, cross-model comparisons challenging.

3) The absence of standardized evaluation benchmarks directly impedes systematic validation of scalability and generalizability. Without unified tasks, datasets, performance metrics, and data encoding protocols, it is exceedingly difficult to rigorously assess whether a quantum Transformer model can generalize beyond the specific, often simplified, conditions under which it was originally tested. This fragmentation hinders reproducibility and the accumulation of reliable evidence for scalable quantum advantage.

\textbf{B. Potential solutions}

Overcoming these critical limitations requires a concerted effort across several fronts. 

1) Establishing unified and rigorous evaluation benchmarks is paramount. This involves proactively developing and integrating diverse and multi-scale datasets, ranging from current small examples to progressively larger and more complex real-world instances. Concurrently, it is essential to standardize evaluation protocols, including clear guidelines for task definitions, performance metrics, data preprocessing pipelines, and, crucially, the selection of competitive classical baselines. Such standardization is indispensable for ensuring transparent, fair, and reproducible comparisons across quantum Transformer implementations.

2) Advancing scalable hybrid quantum-classical architectures is crucial. Research efforts should focus on designing quantum Transformer components that inherently scale more efficiently with increasing problem size, minimizing the growth of quantum resources (qubits, circuit depth, and measurement overhead) while preserving a verifiable quantum advantage. This involves exploring strategies such as hybrid quantum-classical workload balancing \cite{khare2023parallelizing} and adaptive quantum circuit partitioning \cite{daei2020optimized} to optimize resource utilization within NISQ hardware limits. Furthermore, it is imperative to drive theory-driven scalability prediction \cite{larocca2023theory} and develop optimized quantum-classical data pipelines to seamlessly handle larger data volumes.

3) Strengthening empirical validation on large-scale, real-world problems is vital. This necessitates fostering deeper collaboration with industry and domain experts to apply quantum Transformer models to specific, complex real-world tasks \cite{bharti2022noisy}. Such empirical evaluation on larger-scale datasets will expose and help address practical deployment challenges beyond abstract benchmarks. Moreover, researchers must commit to implementing rigorous controlled experiments, conducting stringent comparisons against state-of-the-art classical models to ensure any observed quantum advantage is genuinely compelling.

\subsubsection{Trainability and Optimization Strategy Design} \label{sec4.1.3}

\textbf{A. Current challenges}

In contrast to classical deep learning models, where training techniques such as gradient descent, regularization, and pruning are mature and well-integrated, PQC-based quantum Transformers currently face fundamental challenges in trainability. These challenges go beyond hardware limitations or dataset scale—they reflect the lack of a robust optimization framework that can ensure convergence, stability, and generalization under variational quantum circuits.

One of the most well-documented issues is the barren plateau (BP) phenomenon, in which gradients of the loss function vanish exponentially as the number of qubits or circuit depth increases. This issue is particularly prevalent in deep quantum neural networks, where increasing the number of layers leads to exponentially flatter loss landscapes \cite{mcclean2018barren}. The emergence of barren plateaus can be attributed to several factors, including the presence of global random quantum circuits (leading to concentration of measure effects) \cite{mcclean2018barren}, noise in quantum hardware \cite{wang2021noise}, and the high expressibility or entanglement capacity of the chosen quantum ansatz \cite{larocca2022diagnosing, holmes2022connecting}. In the context of PQC-based Transformers with global token mixing, this becomes especially problematic, as deeper circuits are often required to encode long-range dependencies. In fact, although current PQC-based Transformer models generally use shallow circuits—typically no more than 10 layers—this design choice is itself a reflection of the barren plateau effect: deeper circuits with more parameters have not led to better performance. As a result, most experiments adopt circuit depths that are economically tuned to the threshold where performance saturates. For a comprehensive overview of the barren plateau phenomenon, interested readers are referred to the recent review by Larocca et al. \cite{larocca2025barren}.

Beyond gradient vanishing, PQC training faces an absence of transferable regularization methods. In classical neural networks, techniques such as dropout, weight decay, or early stopping are widely used to prevent overfitting and improve convergence. However, such strategies have no direct counterparts in training quantum models. Dropout, for instance, requires dynamic sub-network masking, which is not natively supported in fixed quantum circuits. Similarly, model pruning is difficult to realize due to the rigid gate structure and parameter entanglement inherent in quantum operations. This lack of robust regularization makes PQC-based models more sensitive to initialization, noise, and circuit design choices.

Furthermore, PQC training encompasses more challenges beyond barren plateaus. The loss landscapes of quantum neural networks can also be plagued by an exponentially large number of local minima \cite{you2021exponentially, anschuetz2022quantum}, which can trap optimization algorithms far from the global optimum and hinder effective learning \cite{zhang2023statistical}. This broader perspective on trainability emphasizes that simply avoiding barren plateaus is insufficient; ensuring convergence to a meaningful solution in a reasonable time also requires navigating these complex landscapes \cite{anschuetz2024unified}.

Trainability evaluation also remains an open issue. Unlike classical models, where metrics such as convergence speed, loss curvature, or generalization gap are routinely reported, quantum models lack standard indicators for training quality. Existing studies rarely report training curves, loss trajectories, or sensitivity analyses, making it difficult to assess reproducibility or identify algorithmic bottlenecks.

Taken together, these issues underscore that trainability in PQC-based quantum Transformers is not merely a matter of circuit design, but a broader challenge of developing a transferable, robust optimization paradigm for quantum learning. The lack of effective regularization, limited adaptability of classical training strategies, and fragile optimization procedures all constrain the practical development and deployment of quantum Transformers.

\textbf{B. Potential solutions}

Addressing the multifaceted trainability challenges in PQC-based quantum Transformers requires a holistic approach, encompassing both theoretical insights and practical algorithmic innovations.

One key strategy focuses on ansatz design to mitigate barren plateaus. While current Q-Transformer models often employ generic shallow circuits, future research can leverage problem-inspired ansatzes that inherently exhibit improved trainability without drastically sacrificing expressive power. Examples include hardware-efficient ansatzes (HVA) \cite{cerezo2021variational}, Scalable Expressive Ansatzes (SEA) \cite{liu2024mitigating}, or finite local-depth circuits (FLDC) which have been shown to avoid barren plateaus under certain conditions \cite{zhang2024absence}. These designs prioritize entanglement and parameter allocation in a way that maintains trainable gradients.

Another crucial aspect is initialization strategies \cite{grant2019initialization}. Proper parameter initialization can significantly impact the trainability of PQCs by placing the optimization algorithm in a region of the loss landscape where gradients are non-vanishing \cite{zhang2022escaping}. Research into more sophisticated initialization methods, tailored to the specific structure of quantum Transformer circuits, is vital.

Furthermore, advancing quantum-aware optimization algorithms is necessary to navigate complex loss landscapes and escape local minima. This could involve developing quantum-inspired optimizers, adaptive learning rate schemes, or hybrid classical-quantum meta-learning approaches that are more robust to noise and the unique topology of quantum cost functions.

Finally, the development of transferable quantum regularization techniques is critical to improve generalization and prevent overfitting. While direct classical counterparts may not exist, exploring quantum analogues or new quantum-specific regularization methods, such as noise-aware training or quantum data augmentation, could enhance the stability and performance of PQC-based Transformers. This overall shift from an architecture-first to a training-aware design paradigm, supported by rigorous algorithmic analysis and standardized benchmarking, will be crucial for the practical advancement of quantum Transformers.

\subsection{Challenges on QLA-based Transformers} \label{subsec4.2}

QLA-based quantum Transformers, which rely on block encoding and QSVT to implement components like attention and feed-forward layers, face a fundamental challenge: model parameters are fixed and not easily trainable. These architectures typically require precompiled unitary matrices, and updating parameters at each training epoch entails reconstructing the entire block-encoding—a process that is computationally intensive and resource-demanding.

This rigidity stands in contrast to PQC-based models, where parameters can be continuously tuned through variational optimization. In QLA-based approaches, however, even slight parameter adjustments require re-synthesizing circuits and managing large ancillary registers, making end-to-end training impractical.

As a result, QLA-based Transformers are currently limited to static inference scenarios. Without efficient mechanisms for parameter updates, they remain theoretical constructs awaiting future breakthroughs in algorithm design or require the integration of PQC modules to become practically usable.

\section{Outlooks}\label{sec6}

As artificial intelligence enters the era of large-scale models, the Transformer architecture has become the foundational backbone for a wide range of applications. In this context, the rise of quantum Transformer research represents not merely a quantum extension of classical architectures, but a potential paradigm shift in model design and computation. Looking ahead, we outline several promising directions for the development of quantum Transformers.

\textbf{A. Toward Quantum Large Models: Embedding Quantum Modules into Scalable Architectures}

The success of large-scale models has been accompanied by unprecedented demands on computational resources. Quantum computing, with its intrinsic parallelism and exponentially large Hilbert space, offers a compelling path toward alleviating these challenges. The diverse array of quantum Transformer variants developed so far can be viewed as a rich toolkit of quantum modules—circuits and subroutines that may be integrated into classical Transformer pipelines.

A natural trajectory is the hybridization of classical large models with quantum components, where quantum circuits are used to replace or enhance parts of attention mechanisms, embedding layers, or positional encodings. Such hybrid architectures may achieve favorable trade-offs between model expressiveness and resource efficiency, particularly in scenarios constrained by power, latency, or memory. Looking further ahead, fully quantum large models operating natively on quantum hardware may become feasible as quantum devices mature, enabling end-to-end quantum-native learning systems.

\textbf{B. Intrinsic Alignment with Physics: Quantum Transformers for Scientific Discovery}

In many areas of science—especially physics, chemistry, and materials science—the data is inherently quantum, either by nature or by measurement. Classical models often require cumbersome preprocessing or approximations to handle such data, potentially losing essential quantum correlations. In contrast, quantum Transformers are uniquely suited to directly process, model, and reason over quantum data.

For instance, in quantum many-body simulations, quantum state reconstruction, or dynamics prediction, quantum Transformers may provide more accurate and compact representations by preserving quantum coherence and entanglement throughout the modeling pipeline. Moreover, tight integration with quantum experimental platforms—such as quantum sensors, simulators, or annealers—can enable quantum Transformers to serve as intelligent control units or real-time decision makers. This convergence opens new frontiers in data-driven scientific discovery.

\textbf{C. From Structural Imitation to Principle-Driven Design: Toward a Theory of Quantum Transformers}

Current quantum Transformer architectures are largely heuristic, often derived by mapping classical structures into quantum analogs. While such designs offer practical starting points, they fall short of harnessing the full theoretical potential of quantum computation. A major future direction lies in reconstructing quantum Transformer design from first principles, guided by quantum information theory, circuit expressivity analysis, and entanglement dynamics.

This effort includes addressing fundamental questions: What class of functions can quantum Transformers efficiently approximate? What are the minimal depth and parameter requirements for a given task? How do barren plateaus, noise, and gradient scaling behave in quantum Transformer training? Furthermore, developing scalable and trainable architectures that preserve performance while remaining robust to decoherence and noise remains a key challenge on the path toward practical deployment.

%%%%%%%%%%%%%%%%%%%%%%%%%%%%%%%%%%%%%%%%%%%%%%%%%%%%%%%%%%%%%%%%%%%%%%%%%%%%%%%%%%%%%%%%%%%%%%%%%%%%%%%%%%%%%

\section{Conclusion}\label{sec7}

This survey has presented the first comprehensive and systematic review of quantum Transformers, offering a detailed landscape of current research from technical architectures to future outlooks. We have categorized existing studies into two primary paradigms, PQC-based and QLA-based approaches, and further proposed a fine-grained taxonomy of PQC-based methods, including QKV-only quantum mapping, quantum pairwise attention (both inner-product and generalized similarity forms), Quantum Holistic Attention, and Quantum-Assisted Optimization. This classification reveals distinct design philosophies underlying their development.

In response to the key questions raised in the introduction—how quantum Transformer architectures are constructed, what distinguishes them, and where quantum advantage may lie—this survey draws several key conclusions.

\textit{How are these quantum Transformer architectures constructed? What are their differences?} - PQC-based models integrate quantum circuits into various components of the Transformer architecture, and typically rely on classical optimizers to update the parameters of PQC during training. Internally, PQC-based approaches further diverge in their design focus: some methods, like QKV-only quantum mapping, primarily quantumize the feature extraction stage; others, such as quantum pairwise attention, target the similarity computation among tokens; while Quantum Holistic Attention methods reformulate the full attention mechanism into a more quantum-friendly structure. In contrast, QLA-based models employ block-encoding and QSVT to implement matrix operations efficiently, but generally do not support parameter updates, resulting in static, inference-oriented architectures.

\textit{Do they demonstrate quantum advantage, and under what conditions?} - Preliminary results suggest that PQC-based models can achieve better performance than classical baselines on small-scale NLP and CV tasks, largely due to their ability to encode data into high-dimensional Hilbert spaces and perform expressive unitary transformations. In scientific domains such as chemistry and high-energy physics, quantum-enhanced models show domain-specific advantages. However, most current gains are limited to constrained scenarios—short sequences, shallow depth, or specialized tasks—rather than broad general-purpose acceleration. For QLA-based models, theoretical quantum advantage manifests as potential exponential speedups in matrix operations.

\textit{Which technical pathways appear most promising?} - At present, research in this field is still in an exploratory stage, making it difficult to determine which approach will ultimately prevail. However, in general, an ideal technological trajectory should aim for higher degrees of quantum integration and greater reductions in classical computational complexity, despite the persistent challenge of trade-offs in quantum resource consumption.

Ultimately, this survey highlights both the promise and the pitfalls of quantum Transformers. While the field remains in its early stage, especially in terms of scaling and generalization, the diverse array of strategies, experimental verifications, and theoretical formulations indicates a vibrant and rapidly evolving research frontier. By synthesizing existing knowledge and identifying both bottlenecks and opportunities, we hope this work serves as a valuable reference for researchers and practitioners aiming to further advance quantum machine learning in the Transformer era.

%%%%%%%%%%%%%%%%%%%%%%%%%%%%%%%%%%%%%%%%%%%%%%%%%%%%%%%%%%%%%%%%%%%%%%%%%%%%%%%%%%%%%%%%%%%%%%%%%%%%%%

\section*{Acknowledgments}

  % % regular IEEE prefers the singular form
  % \section*{Acknowledgment}

This work is funded by the Science and Technology Development Fund, Macau SAR (File No. 0093/2022/A2, 0076/2022/A2, and 0008/2022/AGJ). We sincerely thank Xin Wang, Chenghong Zhu and Guangxi Li for helpful comments during the preparation of this paper.

\section*{Author contributions}

Hui Zhang conceived the structure of the review, conducted the comprehensive literature survey, and was responsible for the initial manuscript drafting.
Qinglin Zhao supervised the entire review process, guided the conceptual framing, and provided in-depth revisions to enhance academic rigor.
Mengchu Zhou offered valuable suggestions on technical classification and contributed to the refinement of the logical structure.
Li Feng reviewed key references and provided insights on organizing the methodological comparisons.
Dusit Niyato provided constructive feedback on the visual presentation of figures and tables, helping improve their clarity and interpretability.
Shenggen Zheng assisted in identifying relevant research trends and contributed to polishing the narrative flow.
Lin Chen reviewed the manuscript thoroughly and provided constructive comments for improving overall coherence and readability.

\section*{Author contributions}

The authors declare no competing interests.

\section*{Materials \& Correspondence}

Correspondence and requests for materials should be addressed to Qinglin Zhao (qlzhao@must.edu.mo).

% \appendices
% \section{Proof of the First Zonklar Equation}
% Appendix one text goes here.

% % you can choose not to have a title for an appendix
% % if you want by leaving the argument blank
% \section{}
% Appendix two text goes here.

% % use section for acknowledgment
% \ifCLASSOPTIONcompsoc
%   % The Computer Society usually uses the plural form

% Can use something like this to put references on a page
% by themselves when using endfloat and the captionsoff option.
\ifCLASSOPTIONcaptionsoff
  \newpage
\fi

% trigger a \newpage just before the given reference
% number - used to balance the columns on the last page
% adjust value as needed - may need to be readjusted if
% the document is modified later
%\IEEEtriggeratref{8}
% The "triggered" command can be changed if desired:
%\IEEEtriggercmd{\enlargethispage{-5in}}

% references section

% can use a bibliography generated by BibTeX as a .bbl file
% BibTeX documentation can be easily obtained at:
% http://mirror.ctan.org/biblio/bibtex/contrib/doc/
% The IEEEtran BibTeX style support page is at:
% http://www.michaelshell.org/tex/ieeetran/bibtex/
%\bibliographystyle{IEEEtran}
% argument is your BibTeX string definitions and bibliography database(s)
%\bibliography{IEEEabrv,../bib/paper}
%
% <OR> manually copy in the resultant .bbl file
% set second argument of \begin to the number of references
% (used to reserve space for the reference number labels box)

\bibliographystyle{IEEEtran}
\bibliography{mybib} % 添加包含引用的数据库文件，如 mybibfile.bib

\end{document}